\documentclass[10pt]{article}

\usepackage{amsmath}
\usepackage{amssymb}

\usepackage{graphicx}

\usepackage{cite}

\usepackage{color} 

\usepackage{subfigure}

\usepackage[labelfont=bf,labelsep=period,justification=raggedright]{caption}

\makeatletter
\renewcommand{\@biblabel}[1]{\quad#1.}
\makeatother

\date{}

\begin{document}

% Title must be 150 characters or less
\begin{flushleft}
{\Large
\textbf{Revisiting Date and Party Hubs: Novel Approaches to Role Assignment in Protein Interaction Networks}
}
% Insert Author names, affiliations and corresponding author email.
\\
Sumeet Agarwal$^{1,2}$, 
Charlotte M. Deane$^{3,4}$, 
Mason A. Porter$^{5,6}$,
Nick S. Jones$^{2,4,6,\ast}$
\\
\bf{1} Systems Biology Doctoral Training Centre, University of Oxford, Oxford OX1 3QD, United Kingdom
\\
\bf{2} Department of Physics, University of Oxford, Oxford OX1 3PU, United Kingdom
\\
\bf{3} Department of Statistics, University of Oxford, Oxford OX1 3TG, United Kingdom
\\
\bf{4} Oxford Centre for Integrative Systems Biology, University of Oxford, Oxford OX1 3QU, United Kingdom
\\
\bf{5} Oxford Centre for Industrial and Applied Mathematics, Mathematical Institute, University of Oxford, Oxford OX1 3LB, United Kingdom
\\
\bf{6} CABDyN Complexity Centre, University of Oxford, Oxford OX1 1HP, United Kingdom
\\
$\ast$ E-mail: Nick.Jones@physics.ox.ac.uk
\end{flushleft}

% Please keep the abstract between 250 and 300 words
\section*{Abstract}
The idea of `date' and `party' hubs has been influential in the
  study of protein-protein interaction networks. Date hubs display low co-expression with their partners, whilst party hubs have
  high co-expression. It was proposed that party hubs are
  local coordinators whereas date hubs are global connectors. 
  Here we show that the reported importance of date hubs to network connectivity can in fact be attributed to a tiny subset of them. 
Crucially, these few, extremely central, hubs do
  not display particularly low expression 
correlation, undermining the idea of a link between this quantity and hub 
function. The date/party distinction was originally motivated by an approximately bimodal distribution of hub
  co-expression; we
  show that this feature is not always robust to
  methodological changes. Additionally, topological properties of hubs do not in general correlate with co-expression. However, we find significant correlations between interaction centrality and the functional similarity of the interacting proteins.
We suggest that thinking in terms of a date/party dichotomy for hubs in protein interaction networks is not meaningful and it might be more useful to conceive of roles for protein-protein interactions rather than for
individual proteins.

% Please keep the Author Summary between 150 and 200 words
% Use first person. PLoS ONE authors please skip this step. 
% Author Summary not valid for PLoS ONE submissions.   
\section*{Author Summary}
Proteins are key components of cellular machinery, and most cellular functions are executed by groups of proteins acting in concert. The study of networks formed by protein interactions can help reveal how the complex functionality of cells emerges from simple biochemistry. Certain proteins have a particularly large number of interaction partners; some have argued that these ``hubs" are essential to biological function. Previous work has suggested that such hubs can be classified into just two varieties: {\it party} hubs, which coordinate a specific cellular process or protein complex; and {\it date} hubs, which link together and convey information between different function-specific modules or complexes. In this study, we re-examine the ideas of date and party hubs from multiple perspectives. By computationally partitioning protein interaction networks into functionally coherent subnetworks, we show that the roles of hubs are more diverse than a binary classification allows. We also show that the position of an interaction in the network is related to the functional similarity of the two interacting proteins: The most important interactions holding the network together appear to be between the most dissimilar proteins. Thus, examining interaction roles may be relevant to understanding the organisation of protein interaction networks.

\section*{Introduction}

Protein interaction networks, constructed from data obtained via techniques such as yeast two-hybrid (Y2H) %%@
screening, do not capture the fact that the actual interactions that occur in vivo depend on prevailing %%@
physiological conditions. For instance, actively expressed proteins vary amongst the tissues in an organism
and also change over time. Thus, the specific parts of the interactome that are active, as well as their %%@
organisational form, might depend a great deal on where and when one examines the network \cite{vid04, tay09}.  %%@
One way to incorporate such information is to use mRNA expression data from microarray experiments. Han {\it et al.} \cite{vid04} 
examined the extent to which hubs in the yeast interactome are co-expressed with %%@
their interaction partners. They defined hubs as proteins with degree at least 5, where 
``degree'' refers to the number of links emanating from a node. Based on the averaged Pearson correlation %%@
coefficient (avPCC) of expression over all partners, they concluded that hubs fall into two distinct classes: %%@
those with a low avPCC (which they called {\it date} hubs) and those with a high avPCC (so-called {\it party} %%@
hubs). They inferred that these two types of hubs play different roles in the modular organisation of the %%@
network: Party hubs are thought to coordinate single functions performed by a group of proteins that are all expressed at the %%@
same time, whereas date hubs are described as higher-level connectors between groups that perform varying functions and are %%@
active at different times or under different conditions. 

The validity of the date/party hub distinction has since been debated in a sequence of papers 
\cite{bat06,vid07,bat07,wil08}, and there appears to be no consensus on the issue. Two established points of %%@
contention are: (1) Is the distribution of hubs truly bimodal (as opposed to exhibiting a continual variation %%@
without clear-cut groupings) and (2) is the date/party distinction that was originally observed a general property of the interactome or an artefact of %%@
the employed data set?  Different statistical tests have %%@
suggested seemingly different answers. However, despite (or in some cases due to) this ongoing debate, the hypothesis has been highly 
prominent in the literature \cite{yu07,kom07,gur08,kim08,vid08,tay09,mis09,kar09,val09,yam09}. Here, following up on the work of Batada {\it et al.} \cite{bat06,bat07}, 
we revisit the initial data and suggest additional problems with the 
statistical methodology that was employed. In accordance with their results, we find that the differing behaviour observed on the %%@
deletion of date and party hubs \cite{vid04}, which seemed to suggest that date hubs were more essential to %%@
global connectivity, was largely due to a very small number of key hubs rather than being a generic property of %%@
the entire set of date hubs. More generally, we use a complementary perspective to Batada {\it et al.} to define 
structural roles for hubs in the context of the modular organisation of protein interaction networks. Our results indicate that there is 
little correlation between expression avPCC and structural roles. In light of this, the more refined categorisation of date, party, and `family' %%@
hubs, which was based on taking into account differences in expression variance in addition to %%@
avPCC \cite{kom07}, also appears inappropriate. A recent study by Taylor {\it et al.} \cite{tay09} argued for the %%@
the existence of `intermodular' and `intramodular' hubs---a categorisation along the same lines as date and %%@
party hubs---in the human interactome. We show that their observation of a binary hub classification is susceptible to 
changes in the algorithm used to normalise microarray expression data or in the kernel function used %%@
to smooth the histogram of the avPCC distribution. The data does not 
in fact display any statistically significant deviation from unimodality as per the DIP test \cite{har85a, har85b}, as has already been observed
by Batada {\it et al.} \cite{bat06, bat07} for yeast data. We revisited the bimodality question because it was a key part of the
original paper \cite{vid04}, and in particular because it made a reappearance in Taylor {\it et al.} \cite{tay09} for human data. However, it is possible that a date-party
like continuum may exist even in the absence of a bimodal distribution,
and this is why we also attempt to examine the more general question of
whether the network roles of hub proteins really are related to their
co-expression properties with interaction partners.

Many real-world networks display some sort of modular organisation, as they can be partitioned into cohesive %%@
groups of nodes that have a relatively high ratio of internal to external connection densities.  Such %%@
sub-networks, known as {\em communities}, often correspond to distinct functional units \cite{gir02, for10, %%@
por09}. Several studies in recent years have considered the existence of community structure in protein-protein %%@
interaction networks \cite{spi03, riv03, yoo04, dun05, gav06, ada06, che06, mar08, lew09}. A myriad of %%@
algorithms have been developed for detecting communities in networks \cite{for10, por09}. For example, the %%@
concept of graph `modularity' can be used to quantify the extent to which the number of links falling within %%@
groups exceeds the number that would be expected in an appropriate random network (e.g., one in which each node %%@
has the same number of links as in the network of interest, but which are randomly placed) \cite{new04}. One of the standard techniques to %%@
detect communities is to partition a network into sub-networks such that graph modularity is maximised %%@
\cite{for10,por09}.

We use the idea of community structure to take a new approach to the problem of hub classification by %%@
attempting to assign roles to hubs purely on the basis of network topology rather than on the basis of %%@
expression data. Our rationale is that the biological roles of date and party hubs are essentially topological %%@
in nature and should thus be identifiable from the network alone (rather than having to be inferred from %%@
additional information). Once we have partitioned the network into a set of 
meaningful communities, it is possible to compute statistics to measure the connectivity of each hub both %%@
within its own community and to other communities. One method for assigning relevant roles to nodes in a %%@
metabolic network was formulated by Guimer\`{a} and Amaral \cite{gui05}, and we follow an analogous procedure %%@
for the hubs in our protein interaction networks. We then examine the extent to which these roles match up with %%@
the date/party hypothesis, finding little evidence to support it.

One might also wonder about the extent to which observed interactome properties are dependent on the particular instantiation of the network being analysed. 
Several papers have discussed at length concerns about the completeness and reliability (or %%@
lack thereof) of existing protein interaction data sets, e.g. \cite{bad02,bad04,hak08,sae08,schw09,ven09,brau09}. Such data have %%@
been gathered using multiple methods, the most prominent of which are Y2H and affinity purification followed by %%@
mass spectrometry (AP/MS). (See the discussion in Materials and Methods.) In a recent paper, Yu {\it et al.} examined %%@
the properties of interaction networks that were derived from different sources, suggesting that experimental %%@
bias might play a key role in determining which properties are observed in a given data set \cite{vid08}. In %%@
particular, their findings suggest that Y2H tends to detect key interactions between protein complexes---so %%@
that Y2H data sets may contain a high proportion of date hubs (i.e., hubs with low partner co-expression)---whereas 
AP/MS tends to detect interactions within complexes, so that hubs in AP/MS-derived networks are %%@
predominantly highly co-expressed with their partners (i.e., these networks will contain party hubs). This indicates that a %%@
possible reason for observing the bimodal hub avPCC distribution \cite{vid04} is that the interaction data sets used information that 
was combined from both of these sources. Here we compare several yeast interaction data sets and find both widely differing structural properties and a surprisingly low level of overlap.

Finally, as an alternative to the node-based date/party categorisation, we suggest thinking about topological roles in networks by defining 
measures on links rather than on nodes. In other words, one can attempt to categorise interactions between proteins rather than the proteins %%@
themselves. We use a well-known measure of link significance known as {\em betweenness centrality} %%@
\cite{fre77,gir02} and examine its relation to phenomena such as protein co-expression and functional overlap. %%@
Here as well we find little evidence of a significant correlation with expression PCC of the interactors. However, there seems to %%@
be a reasonably strong relation between link betweenness and functional similarity of the interacting proteins, so that
link-centric role definitions might have some utility.

In summary, we have examined the proposed division of hubs in the protein interaction network into the date and %%@
party categories from several different angles, demonstrating that prior arguments in favour of a date/party %%@
dichotomy appear to be susceptible to various kinds of changes in the data and methods used. 
Observed differences in network vulnerability to attacks on the two hub types seem to arise from only a small number of particularly important hubs. 
These results strengthen the existing evidence against the existence of date and party hubs. Furthermore, a detailed analysis of 
network topology, employing the novel perspective of community structure and the roles of hubs within this context, suggests that the picture is more 
complicated than a simple dichotomy. Proteins in %%@
the interactome show a variety of topological characteristics that appear to lie along a continuum---and there %%@
does not exist a clear correlation between their location on this continuum and
the avPCC of expression of their interaction partners. On the other hand, investigating link (interaction) %%@
betweenness centralities reveals an interesting relation to 
the functional linkage of proteins, suggesting that a framework incorporating a more nuanced notion of roles %%@
for both nodes and links might provide a better framework for understanding the organisation of the %%@
interactome.

% Results and Discussion can be combined.
\section*{Results}

\subsection*{Revisiting Date and Party Hubs}

The definitions of date and party hubs are based on the expression correlations of hubs with their interactors 
in the protein interaction network . Specifically, the avPCC has been computed for each hub and its distribution was %%@
observed by Han {\it et al.} \cite{vid04} to be bimodal in some cases. A date/party threshold value of avPCC (for a given expression data set) was 
defined in order to optimally separate the two types of hubs \cite{vid04}.

We have re-examined the data sets and analyses that were used to propose the existence and dichotomy of date %%@
versus party hubs. In the original studies on yeast data \cite{vid04, vid07}, any hub that exhibited a %%@
sufficiently high avPCC (i.e., any hub lying above the date/party threshold) on {\em any one} expression data set was 
identified as a party hub. Batada {\it et al.} \cite{bat07} noted that this definition causes the date/party
assignment to be overly conservative, in that a hub's status is unlikely to change as a result of additional expression data.
In fact, some of the original expression data sets were quite small, containing fewer than 10 data %%@
points per gene. This suggests that classification of proteins as `party' hubs was based on high co-expression with %%@
partners for just a small number of conditions in a single microarray experiment, even though such %%@
co-expression need not have been observed in other conditions and experiments. For instance, Han {\it et al.} found %%@
108 party hubs in their initial study \cite{vid04}. However, calculating avPCC across their entire expression %%@
compendium (rather than separately for the five constituent microarray data sets) and using the date/party %%@
threshold specified by the authors for this compendium avPCC distribution yields just 59 party hubs. Using only %%@
the ``stress response" data set \cite{gas00}, which comprises over half of the data points in their compendium %%@
and is substantially larger than the other 4 sets, yields 74 party hubs. Thus, the results of applying this %%@
method to categorise hubs depend heavily on the expression data sets that one employs and is vulnerable to %%@
variability in smaller microarray experiments. 

\begin{figure}[!ht]
\begin{center}
\includegraphics{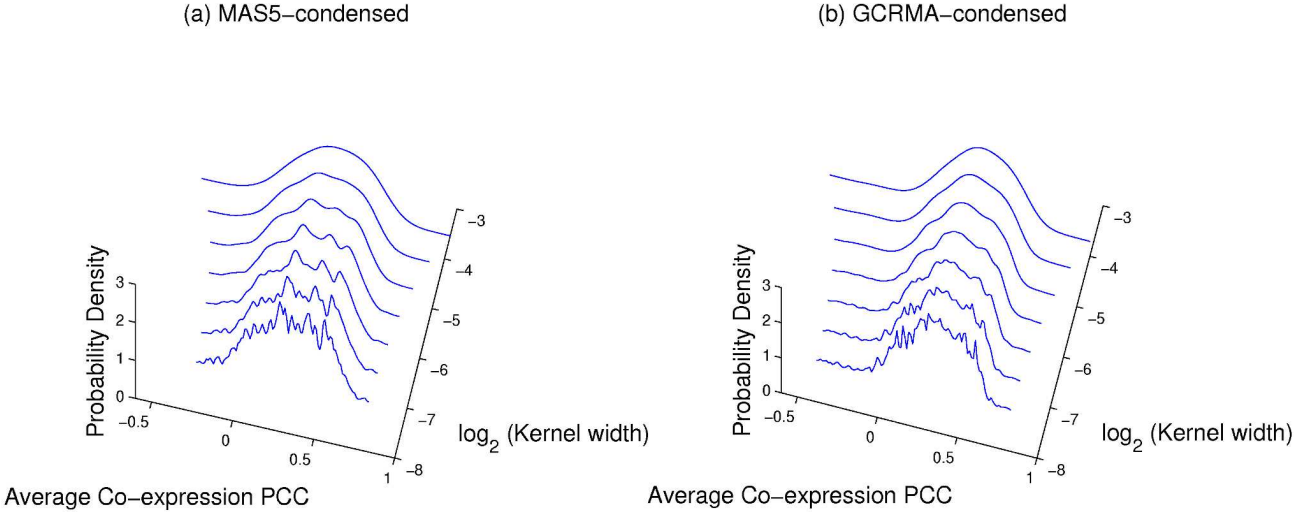}
\end{center}
\caption{{\bf Variation in hub avPCC distribution.} Probability density plots of the distribution of hub avPCC values for human interaction data from %%@
OPHID (provided by Taylor {\it et al.} \cite{tay09}). Gene expression data from GeneAtlas \cite{su04}, normalised %%@
using (a) MAS5 and (b) GCRMA \cite{lim07}. Curves obtained using a normal smoothing kernel function at varying %%@
window widths. Hartigan's DIP test for unimodality \cite{har85a, har85b} returns values of 0.0087 ($p$-value $\approx 0.821$) for (a) and 0.0062 ($p$-value $\approx 0.998$) for (b), 
indicating no significant deviation from unimodality in either case.} 
\label{fig:tay}
\end{figure}

Recent support for the idea of date and party hubs appeared in a paper that considered data relating to the %%@
human interactome; the authors found multimodal distributions of avPCC values, seemingly supporting a binary hub classification \cite{tay09}.
We used an interaction data set provided by Taylor {\it et al.} \cite{tay09} (an updated version of the one %%@
used in their paper, sourced from the Online Predicted Human Interaction 
Database (OPHID) \cite{bro05}; see Materials and Methods), and found that the form of the distribution of hub
avPCC that they observed is not robust to methodological changes. For instance, raw intensity data from %%@
microarray probes has to be processed and normalised in order to obtain comparable expression values for each %%@
gene \cite{lim07}. The expression data used by Taylor {\it et al.} \cite{tay09} (taken from the human GeneAtlas \cite{su04}) %%@
was normalised using the Affymetrix MAS5 algorithm \cite{hub02}; when we repeated the analysis using the same data %%@
normalised by the GCRMA algorithm \cite{wu04} (which is the preferred method to control for probe affinity) instead of by MAS5,
we obtained significantly different results. Figure %%@
\ref{fig:tay} depicts the avPCC distributions for hubs (defined as the top 15\% %%@
of nodes by degree \cite{tay09}, corresponding in this case to degree 15 or greater) in the two cases. We obtained density 
plots for varying smoothing kernel widths. The GCRMA-processed data does not appear to lead to a %%@
substantially bimodal distribution at any kernel width, whereas the MAS5-processed data appears to give %%@
bimodality for only a relatively narrow range of widths and could just as easily be regarded as trimodal. We also 
used Hartigan's DIP test \cite{har85a,har85b,dipcode} to check whether either of the two versions of the expression data gives a distribution of avPCC values 
showing significant evidence of bimodality. The DIP value is a measure of how far an observed distribution deviates from the best-fit unimodal distribution, 
with a value of 0 corresponding to no deviation. We used a bootstrap sample of 10,000 to obtain $p$-values for the DIP statistic. We found no significant 
deviation from unimodality: for MAS5, the DIP value is $0.0087$ ($p$-value $\approx 0.821$) and for GCRMA the DIP value is $0.0062$ ($p$-value $\approx 0.998$). This suggests that the
apparent bimodal or trimodal nature of some of the curves in Figure \ref{fig:tay} is illusory and not statistically robust.

We also find variability across different interaction data sets: For instance, we analysed the recent %%@
protein-fragment complementation assay (PCA) data set \cite{tar08} and found no clear evidence of a bimodal %%@
distribution of hubs along date/party lines (data not shown). Even in cases where multimodality is observed, it might %%@
be arising as a consequence or artefact of combining different types of interaction data; there are believed to be %%@
significant and systematic biases in which types of interactions each data-gathering method is able to obtain %%@
\cite{vid08,tar08,lew09}. For instance, analysing avPCC values on the stress response expression data set \cite{gas00}
for hubs in networks obtained from Y2H or AP/MS alone \cite{vid08}, we find that 100\% (259/259) are date hubs in the 
former but that only about 30\% (56/186) are date hubs in the latter. At the moment it is reasonable to entertain the 
possibility that new kinds of interaction tests might smear out the observed bimodality; this appears to be the case 
with the PCA data set.

\begin{figure}[!ht]
\begin{center}
\includegraphics{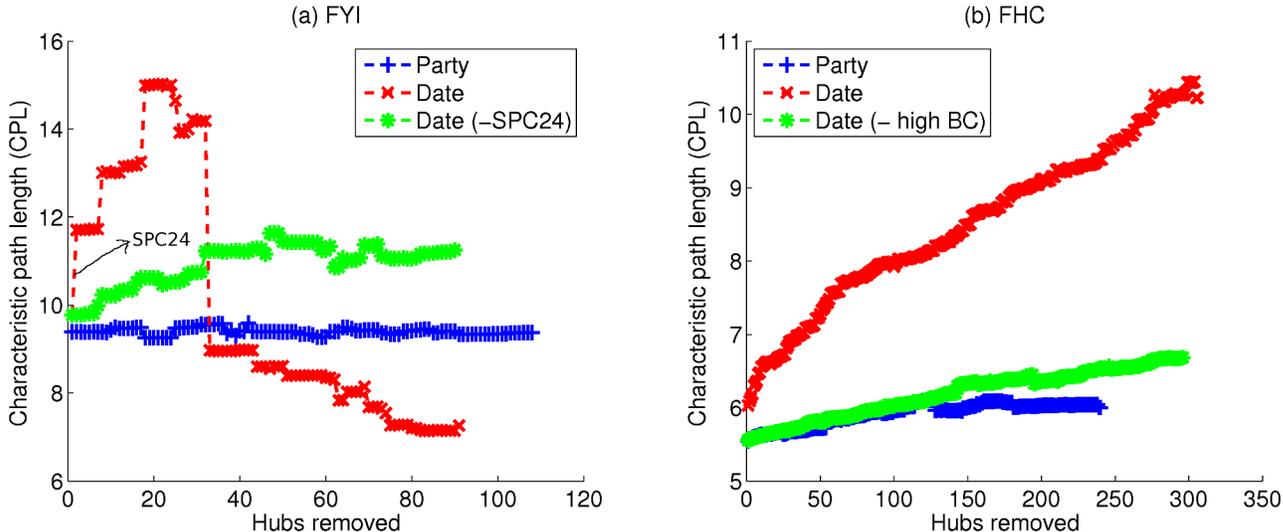}
\end{center}
\caption{{\bf Effects of hub deletion on network connectivity.} (a) FYI network \cite{vid04}. `Date ($-$ SPC24)' refers to the
set of date hubs minus the protein SPC24. In each case, we used the complete network 
consisting of 1379 nodes as the starting point and then deleted all hubs in the given set from the network in %%@
order of decreasing degree. The characteristic path length is the mean of the lengths of all finite paths %%@
between two nodes in the network. (b) FHC network \cite{vid07}. `Date ($-$ high BC)' refers to the set of date hubs minus the 10 hubs with the highest %%@
betweenness centrality (BC) values (listed in Table \ref{tab:hubdel}). We used the upper bound on the BC for party hubs as a threshold to define %%@
these 10 `high BC' date hubs. (Note: Results similar to those presented here are obtained if the hubs are divided into bottleneck/non-bottleneck 
categories \cite{yu07} instead of date/party categories.)
}
\label{fig:hubdel}
\end{figure}

One of the key pieces of evidence used to argue that date and party hubs have distinct topological properties %%@
was the apparent observation of different effects upon their deletion from the network. Removing date hubs %%@
seemed to lead to very rapid disintegration into multiple 
components, whereas removal of party hubs had much less effect on global connectivity \cite{vid04, vid07}.  %%@
However, it has been observed that removing just the top 2\% of hubs by degree from the comparison of deletion %%@
effects obviates this difference, suggesting 
that the observation is actually due to just a few extreme date hubs \cite{bat07}. In order to study this in %%@
greater detail, and to isolate the extreme hubs, we used node betweenness centrality \cite{fre77} (see Materials and Methods), a standard metric %%@
of a node's importance to network connectivity (this need not be strongly correlated with degree). We found that in the 
original `filtered yeast interactome' (FYI) data set \cite{vid04}, date hubs have on average somewhat higher %%@
betweenness centralities ($1.79 \times 10^4$ for 91 date hubs versus $1.07 \times 10^4$ for 108 party hubs, two-sample $t$-test $p$-value $\approx 0.08$).
However, there happens to be one date hub (SPC24/UniProtKB:Q04477, a highly connected protein involved in chromosome segregation %%@
\cite{uni08}) that has an exceptionally high betweenness ($2.45 \times 10^5$) in this network. When the set of date hubs minus %%@
this one hub is targeted for deletion, we find that the observed difference between date and party hubs is greatly reduced %%@
(Figure \ref{fig:hubdel}(a)).

It was subsequently shown that the FYI network was particularly sparse; as more data became available, the %%@
updated filtered high-confidence (FHC) data set was used to perform the same analysis \cite{vid07} (we also looked at the 
Y2H-only and AP/MS-only networks \cite{vid08}; see Figure S1). In the case of FHC, the network 
did not break down on removing date hubs but nevertheless displayed a substantially greater increase in characteristic path %%@
length (CPL) than seen for party hub deletion. For FHC too, date hubs have, on average, higher betweenness values than party hubs ($3.7 %%@
\times 10^4$ for 306 date hubs versus $2.15 \times 10^4$ for 240 party hubs, $p$-value $\approx 0.06$). However, the larger average is %%@
due almost entirely to a small number of hubs with unusually high betweennesses, as removing the top 10 date hubs %%@
by betweenness (which all had values higher than any party hub) greatly reduced the difference between the distributions ($p$-value $\approx 0.29$). Furthermore, the %%@
removal of just these 10 hubs from the set of targeted date hubs is sufficient to virtually obviate the %%@
difference with party hubs, as shown in Figure \ref{fig:hubdel}(b). Notably, the set of 10 high-betweenness hubs %%@
includes prominent proteins such as Actin (ACT1/UniProtKB:P60010), Calmodulin (CMD1/UniProtKB:P06787), and the 
TATA binding protein (SPT15/UniProtKB:P13393), which are known to be key to important cellular processes (Table \ref{tab:hubdel}).
Thus, we can account for the critical nodes for network connectivity using just a few major hubs, and most of %%@
the proteins that are classified as date hubs appear to be no more central than the party hubs. High betweenness nodes have 
previously been referred to as {\it bottlenecks} \cite{yu07} and it has been suggested that these are in general highly central 
and tend to correspond to date hubs. However, the same sort of analysis on the Yu {\it et al.} data set \cite{yu07} once again 
revealed that only the top 0.5\% or so of nodes by betweenness are truly critical for connectivity (data not shown). Additionally, 
the 10 key hubs in the FHC network show a wide range of avPCC values (Table \ref{tab:hubdel}): high betweenness does not 
necessitate low avPCC. Similarly, we found no strong correspondence between bottleneck/non-bottleneck and date/party distinctions
across multiple data sets. These observations further weaken the claim that there is an inverse relation between a hub's avPCC 
and its central role in the network. 

\begin{table}[!ht]
\caption{\bf{High-betweenness Hubs in the FHC Network}} 
\begin{tabular}{llllll}
\hline
{\bf Protein} & {\bf UniProtKB} & {\bf Degree} & {\bf AvPCC} & {\bf BC($/10^5$)} & {\bf Functions}\\
\hline
CDC28  & P00546 & 202  &  0.06 & 19.99 & Essential for the completion of the start, the\\ 
&&&&& controlling event, in the cell cycle\\ 

RPO21 &	P04050 & 58  &  0.05 & 3.56 & Catalyses the transcription of DNA into RNA\\

SMT3 & Q12306 & 42  &  0.08 & 3.07 & Not known; suppressor of MIF2\\ 
&&&&& (UniProtKB:P35201) mutations\\

ACT1  & P60010 & 35  &  0.13 & 2.83 & Cell motility\\

HSP82 & P02829 & 37  &  0.19 & 2.51 & Maturation, maintenance, and regulation\\
&&&&& of proteins involved in cell-cycle\\
&&&&& control and signal transduction\\

SPT15 & P13393 & 50  &  0.12 & 2.45 & Regulation of gene expression\\
&&&&& by RNA polymerase II\\

CMD1  & P06787 & 46  &  0.05 & 2.11 & Mediates the control of a large number of\\
&&&&& enzymes and other proteins\\

PAB1  & P04147 & 25  &  0.28 & 1.92 & Important mediator of the roles of the\\
&&&&& poly(A) tail in mRNA biogenesis,\\ 
&&&&& stability, and translation\\

PSE1  & P32337 & 24   &  0.28 & 1.73 & Nuclear import of ribosomal\\
&&&&& proteins and protein secretion\\

GLC7 & P32598 & 35 &  -0.01 & 1.55 & Glycogen metabolism, meiosis, translation,\\
&&&&& chromosome segregation, cell polarity,\\ 
&&&&& and cell cycle progression\\
\hline
\end{tabular}
\begin{flushleft}
List of the 10 high-betweenness hubs in the FHC network 
\cite{vid07}, with UniProtKB accessions \cite{uni08}, degrees, avPCC values (as %%@
computed using the `Compendium' expression data set \cite{vid04,kem02}), betweenness centrality (BC) values, and selected functional annotations from %%@
UniProtKB.
\end{flushleft}
\label{tab:hubdel}
\end{table}

\subsection*{Topological Properties and Node Roles}

\begin{figure}[!ht]
\begin{center}
\includegraphics{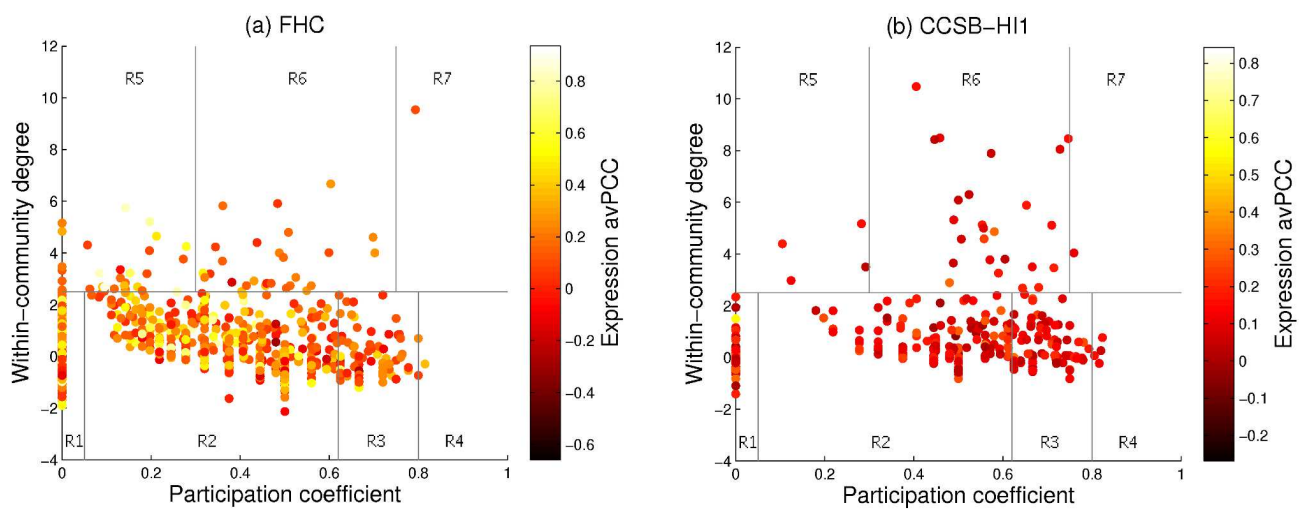}
\end{center}
\caption{{\bf Topological node role assignments and relation with avPCC.} Plots for (a) Yeast network 
(FHC \cite{vid07}---2,233 nodes, 63 communities) and (b) Human network (CCSB-HI1 \cite{vid05}---1,307 nodes, 38 communities) 
(see Materials and Methods for details). Following Guimer\`{a} and Amaral \cite{gui05}, we designate the roles %%@
as follows: R1 -- Ultra-peripheral; R2 -- Peripheral; R3 -- Non-hub connector; R4 -- Non-hub kinless; R5 -- %%@
Provincial hub; R6 -- Connector hub; and R7 -- Kinless hub. We colour proteins according to the avPCC of %%@
expression with their interaction partners. We computed expression avPCC using the stress response data set %%@
\cite{gas00} (which was the largest, by a considerable margin, of the expression data sets used in the original study \cite{vid04}) 
for FHC and COXPRESdb \cite{oba08} for CCSB-HI1. No partner expression data was available for a few proteins (25 in FHC, 1 in 
CCSB-HI1)---these are not shown on the plots.
}
\label{fig:fcpcc}
\end{figure}

In principle, one should be able to view a categorisation of hubs according to the date/party dichotomy %%@
directly in the network structure, as the two kinds of hubs are posited to have 
different neighbourhood topologies. We thus leave gene expression data to one side for the moment and focus on what can
be inferred about node roles purely from network topology. Guimer\`{a} and Amaral \cite{gui05} have proposed a %%@
scheme for classifying nodes into topological roles in a modular network according to their pattern of %%@
intramodule and intermodule connections. Their classification uses two statistics for each node---within-community degree 
and participation coefficient (a measure of how well spread out a node's links are amongst all communities, including its own)---and 
divides the plane that they define into regions encompassing seven possible roles (see Materials and Methods for details). 
We depict these regions in Figure \ref{fig:fcpcc}, which shows the node roles for yeast (FHC \cite{vid07}) and human 
(Center for Cancer Systems Biology Human Interactome version 1 (CCSB-HI1) \cite{vid05}) 
data sets, which we computed based on communities detected by optimising modularity via the Potts method \cite{rei06} (see Text S1 for details, and Figure S4 and Table S1 for indications of the structural and functional coherence of the communities, respectively). 
Also, when partitioning the network using this method, one can adjust the resolution to get more or fewer communities. In Figure S2, we 
show the results of this computation repeated for two other values of the resolution parameter. In each case, we obtain a similar pattern to the results 
shown here, and the conclusions below are valid across the multiple resolutions examined.

\begin{figure}[!ht]
\begin{center}
\includegraphics{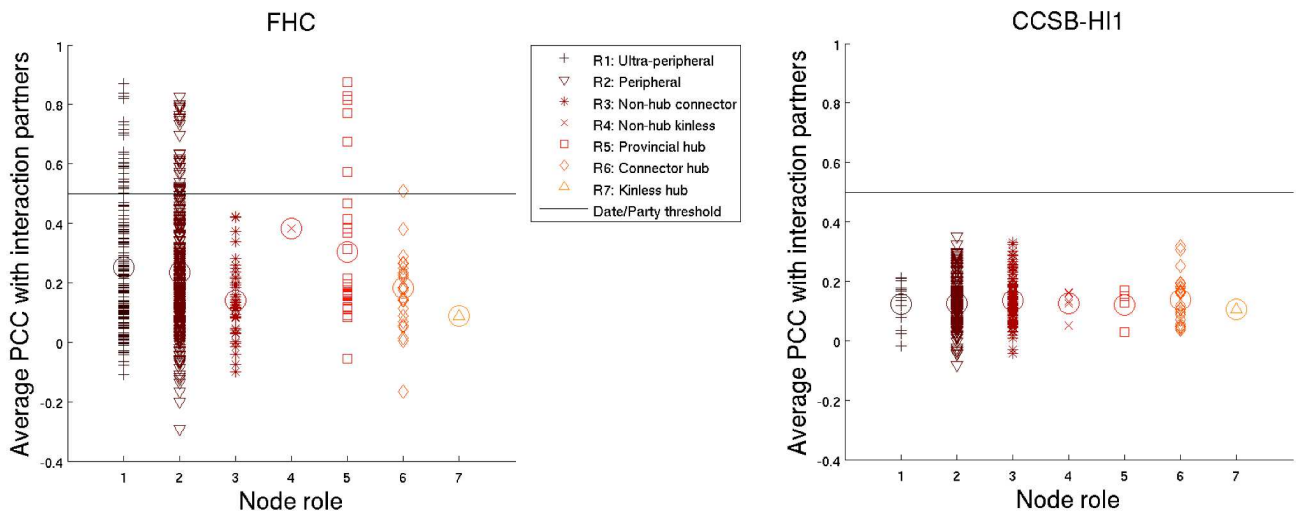}
\end{center}
\caption{{\bf Rolewise hub avPCC distributions.} Plots show node role versus average expression correlation 
with partners for hubs in yeast (FHC \cite{vid07}---553 hubs with a minimum degree of 7) and human (CCSB-HI1 \cite{vid05}---326 hubs with a minimum 
degree of 4) networks. Larger circles represent means over all nodes in a given role. Note that `hub' as used in the role 
names refers only to within-community hubs, but all of the depicted nodes 
are hubs in the sense that they have high degree. In each case, we determined the degree threshold so that %%@
approximately the top 20\% highest-degree nodes are considered to be hubs. We also fixed the date/party avPCC %%@
threshold at 0.5, in accordance with Bertin {\it et al.} \cite{vid07}.
}
\label{fig:fcdp}
\end{figure}

Some of the topological roles defined by this method correspond at least to some extent to those ascribed to %%@
date/party hubs.  For instance, one might argue that party hubs ought to be `provincial hubs', which have many links within their %%@
community but few or none outside. Date hubs might be construed as `non-hub connectors' or `connector hubs', %%@
both of which have links to several different modules; they could also fall into the `kinless' roles (though %%@
very few nodes are actually classified as such). We thus sought to examine the relationship between the %%@
date/party classification and this topological role classification. In Figure \ref{fig:fcpcc}, we colour %%@
proteins according to their avPCC. In Figure \ref{fig:fcdp}, we present the same data in a more compact form, %%@
as we only show the hubs (defined as the top 20\% of nodes ranked by degree \cite{vid07}) in the two %%@
interaction networks, plotting them according to node role and avPCC. The horizontal lines correspond to an %%@
avPCC of 0.5, which was the threshold used to distinguish date and party hubs in the yeast interactome %%@
\cite{vid07}.

One immediate observation from these results is that the avPCC threshold clearly does not carry over to the %%@
human data. In fact, all of the hubs in the latter have an avPCC of well below 0.5. Even if we utilise a %%@
different threshold in the human network, we find that there is little difference in the avPCC distribution %%@
across the topological roles, suggesting that no meaningful date/party categorisation can be made (at least for %%@
this data set). This might be the case because the human data set represents only a small fraction of %%@
the actual interactome. Additionally, it is derived from only one technique (Y2H) and is thus not %%@
multiply-verified like the yeast data set. 

For yeast, we see that hubs below the threshold line (i.e., the supposed date hubs) include not only virtually all of %%@
those that fall into the `connector' roles but also many of the `provincial hubs'.  On the other hand, those %%@
that lie above the line (i.e., the supposed party hubs) include mainly the provincial hub and peripheral categories.  %%@
Although one can discern a difference in role distributions above and below the threshold, it is not very %%@
clear-cut and the so-called date hubs fall into all 7 roles.  It would thus appear that even for yeast, the %%@
distribution of hubs does not clearly fall into two types (the original statistical analysis has already been disputed by Batada et %%@
al. \cite{bat06, bat07}), and the properties attributed to date and party hubs \cite{vid04} do not seem to %%@
correspond very well with the actual topological roles that we estimate here. Indeed, these roles are more %%@
diverse than what can be explained using a simple dichotomy.

\subsection*{Data Incompleteness and Experimental Limitations}

It has been proposed that date and party hubs play different roles with respect to the modular structure of protein 
interaction data. As there are diverse examples of such data, one might ask to what extent entities like date and %%@
party hubs can be consistently defined across these. In order to investigate the extent of network overlap and the 
preservation of the interactome's structural properties (such as community structure and node roles) for different 
data sets and data-gathering techniques, we compared statistics and results for four different yeast interaction data sets: 
FYI, FHC, Database of Interacting Proteins core (DIPc), and PCA (see Table \ref{tab:ds} and Materials and Methods for details of these).  
Our motivation for these choices of data sets (aside from PCA) was that they all encompass multiply-verified or high-confidence %%@
interactions. We also used PCA data because it is from the first large-scale screen with a new technique that %%@
records interactions in their natural cellular environment \cite{tar08}.
For each data set, we counted the number of nodes and links in common using pairwise comparisons in the largest %%@
connected component of the network. For the overlapping portions, we then computed the extent of overlap in %%@
node roles and communities.  For the latter, we employed the Jaccard distance \cite{jac01}, which ranges from 0 for %%@
identical partitions to 1 for entirely distinct ones (see Materials and Methods). In Table \ref{tab:comp},
we present the results of our binary comparisons of %%@
the yeast data sets.

\begin{table}[!ht]
\caption{\bf{Protein Interaction Data Sets}}
\begin{tabular}{lllllll}
\hline
\bf{Data set name} & \bf{Species} & \multicolumn{2}{l}{\bf{Nodes}} & \multicolumn{2}{l}{\bf{Links}} &
\bf{Source} \\
&& Total & LCC & Total & LCC & \\
\hline
Online Predicted Human & {\it H. sapiens} & 8,199 & 7,984 & 37,968 & 37,900 & Brown \& Jurisica \cite{bro05} \\
Interaction Database (OPHID) &&&&&& (curated by Taylor {\it et al.} \cite{tay09}) \\

Filtered yeast  & {\it S. cerevisiae} & 1379 & 778 & 2493 & 1798 & Han {\it et al.} \cite{vid04} \\
interactome (FYI) &&&&&& \\ 

Filtered high- & {\it S. cerevisiae} & 2559 & 2233 & 5991 & 5750 & Bertin {\it et al.} \cite{vid07} \\
confidence (FHC) &&&&&& \\

Database of Interacting & {\it S. cerevisiae} & 2808 & 2587 & 6212 & 6094 & \emph{http://dip.doe-mbi.ucla.edu/} \\
Proteins core (DIPc) &&&&&& (October 2007 version) \\

Center for Cancer Systems & {\it H. sapiens} & 1,549 & 1,307 & 2,611 & 2,483 & Rual {\it et al.} \cite{vid05} \\
Biology Human Interactome &&&&&&\\
version 1 (CCSB-HI1) &&&&&&\\

Protein-fragment & {\it S. cerevisiae} & 1124 & 889 & 2770 & 2407 & Tarassov {\it et al.} \cite{tar08} \\
complementation assay (PCA) &&&&&&\\
\hline
\end{tabular}
\begin{flushleft}
The protein interaction data sets that we used in this paper. LCC refers to the largest connected
component.
\end{flushleft}
\label{tab:ds}
\end{table}

\begin{table}[!ht]
\caption{\bf{Comparisons of Yeast Data Sets}}
{\small
\begin{tabular}{lllll}
\hline
\bf{Data sets} & \bf{Common}  & \bf{Links in}  & \bf{Between-community} & \bf{Role$^3$} \\
\bf{(number of nodes)} & \bf{nodes$^1$} & \bf{overlap$^2$} & \bf{Jaccard distance$^3$} &
\bf{overlap$^4$} \\
\hline
FYI (778) vs. FHC (2233) & 714 & FYI--1444; FHC--2027; Both--1195 & 0.76 & 332 (47\%) \\
FYI (778) vs. DIPc (2587) & 660 & FYI--1310; DIPc--1698; Both--956 & 0.77 & 265 (40\%) \\
FHC (2233) vs. DIPc (2587) & 1661 & FHC--4395; DIPc--4141; Both--2665 & 0.85 & 854 (51\%) \\
FYI (778) vs. PCA (889) & 165 & FYI--154; PCA--180; Both--65 & 0.74 & 109 (66\%) \\
FHC (2233) vs. PCA (889) & 460 & FHC--512; PCA--667; Both--187 & 0.86 & 214 (47\%) \\
DIPc (2587) vs. PCA (889) & 492 & DIPc--568; PCA--782; Both--183 & 0.86 & 206 (42\%) \\
\hline
\end{tabular}
}
\begin{flushleft}
Pairwise comparisons of the largest connected components of different yeast protein interaction %%@
data sets. Notes:
{\bf 1} Proteins occurring in both networks.
{\bf 2} Links amongst the common nodes as counted in the previous column: individually in either network and %%@
common to both networks.
{\bf 3} Communities and node roles computed over entire data sets; for pairwise comparison, we then narrow down communities in each case
to only those nodes also present in the data set being compared to.
{\bf 4} The number of nodes with the same role classification in both networks, and their percentage as a share of the entire set of common nodes. 
\end{flushleft}
\label{tab:comp}
\end{table}

Table \ref{tab:comp} reveals that there are large variations amongst the different networks reported in the %%@
literature. FYI, FHC, and DIPc are all regarded as high-quality data sets, yet they contain numerous %%@
disparate interactions. PCA has a very low overlap with 
both FYI and DIPc (considered separately), suggesting that it provides data that is not captured by either Y2H or AP/MS screens. Such %%@
differences unsurprisingly lead to nodes having variable community structure between data sets. The Jaccard distance for each pairwise %%@
comparison amongst the 4 networks is around 0.8, so on %%@
average the intersection of communities for the same node covers only about a fifth of their union (for comparison purposes, communities are 
computed over the complete network in each case, and then each community is pruned to 
retain only those nodes also present in the other network). Because we compute topological node roles relative to community structure, 
it is not surprising that the role overlap is also not very high in any of the cases. 

Given the above, it is difficult to make any general inferences regarding proteome organisation from results on %%@
existing protein interaction networks. They depend a great deal on the explored data set, which in each case %%@
represents only part of the total interactome and may also contain substantial noise. 

\subsection*{The Roles of Interactions}

\begin{figure}[!ht]
\begin{center}
\includegraphics{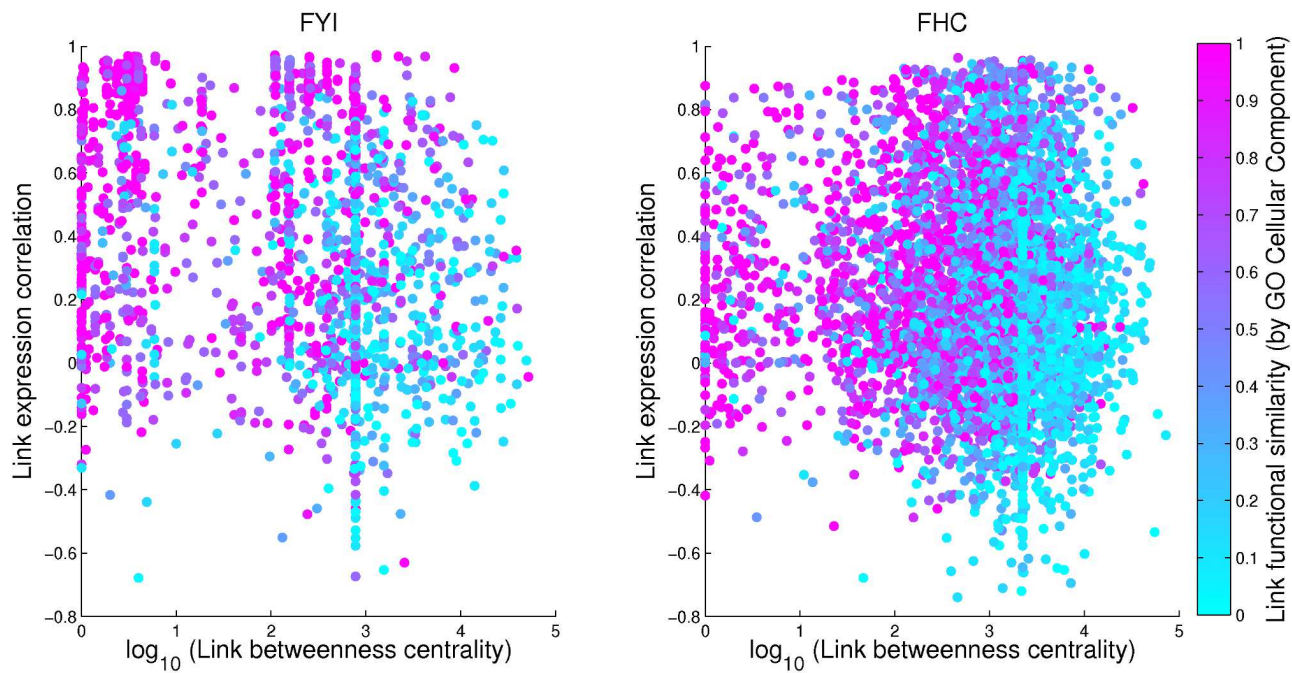}
\end{center}
\caption{{\bf Relating interaction betweenness, co-expression, and functional similarity.}
Plots show link betweenness centralities versus expression correlations, with points coloured according to average  
similarity of interactors' GO (Cellular Component) annotations, for two protein interaction data sets: FYI \cite{vid04}
(778 nodes, 1,798 links) and FHC \cite{vid07} (2,233 nodes, 5,750 links). PCC values of log(link betweenness) with functional similarity are $-0.51$ ($z$-score $\approx -23.9$, $p$-value $\approx 1.4 \times 10^{-126}$) for FYI, $-0.46$ ($z$-score $\approx -37.2$, $p$-value $\approx 1.6 \times 10^{-303}$) for FHC.
}
\label{fig:simbce}
\end{figure}

Most research on interactome properties has focused on node-centric metrics, which draws on the perspective of %%@
individual proteins (e.g., \cite{jeo01,vid04,kom07,zot08}). Here we try an alternative approach that instead uses link-centric metrics in order to %%@
examine how the topological properties of interactions in the network relate to their function. In order to %%@
quantify the importance of a given link to global network connectivity, we use link betweenness centrality \cite{gir02,fre77} 
(see Materials and Methods). We investigate the relationship between link %%@
betweenness and the expression correlation for a given interaction.  If date and party hubs genuinely exist, %%@
one might expect a similar sort of dichotomy for interactions, with more central interactions having lower %%@
expression correlations and vice versa. That is, given the hypothesised functional roles of date and party %%@
hubs, most intermodular interactions would connect to a date hub, whereas most intramodular %%@
interactions would connect to a party hub. In Figure \ref{fig:simbce}, we depict all of the interactions in two %%@
yeast data sets, which we position on a plane based on the values of their link 
betweenness and interactor expression PCC (calculated using the stress response data set as before). %%@
Additionally, we colour each point according to the level of functional similarity between the interacting %%@
proteins, as determined by overlap in GO (Cellular Component) annotations (see Materials and Methods). We also obtain similar results 
using the other two GO ontologies, which are shown in Figure S3.

For the FHC data set, we find no substantial relation between expression PCC and the logarithm of link %%@
betweenness (linear Pearson correlation $\approx -0.04$, $z$-score $\approx -3.1$, $p$-value $\approx 0.0022$). For the FYI data set, there is 
a larger correlation ($-0.31$, $z$-score $\approx -13.6$, $p$-value $\approx 4.5 \times 10^{-42}$). Correspondingly,
we observe a dense cluster of interactions in the top left (i.e., they have low %%@
betweennesses and high expression correlations), but most of these are interactions within ribosomal complexes. If %%@
one removes such interactions from the data set, then here too one finds only a small correlation ($-0.12$, $z$-score $\approx -4.5$, 
$p$-value $\approx 5.8 \times 10^{-6}$) between expression PCC and (log of) link betweenness. (Note that ribosomal proteins were already removed from 
FHC \cite{vid07}.) On the other hand, we find a fairly strong correlation between link betweenness (on a log-scale) and similarity in cellular %%@
component annotations (which can be used as a measure of co-localisation): the PCC values are $-0.51$ ($z$-score $\approx -23.9$, 
$p$-value $\approx 1.4 \times 10^{-126}$) for FYI and $-0.46$ ($z$-score $\approx -37.2$, $p$-value $\approx 1.6 \times 10^{-303}$) for FHC 
(very similar values are obtained for the Spearman rank correlation coefficient: 
$-0.52$ for FYI and $-0.47$ for FHC). In particular, there appears to be a natural threshold at the modal value %%@
of betweenness. (As discussed in Materials and Methods, this is a finite-size effect.) This is somewhat %%@
reminiscent of the weak/strong tie distinction in social networks \cite{rap57, gra73}, as the `weak' (high %%@
betweenness) interactions serve to connect and transmit information between distinct cellular modules, which are %%@
composed predominantly of `strong' (low betweenness) interactions. For instance, we found that interactions %%@
involving kinases fall largely into the `weak' category. Additionally, GO terms such as 
intracellular protein transport, GTP binding, and nucleotide binding were enriched significantly in proteins %%@
involved in high-betweenness interactions.

\section*{Discussion}

In this paper, we have analysed modular organisation and the roles of hubs in protein interaction
networks. We revisited the possibility of a date/party hub dichotomy and found points of concern. %%@
In particular, claims of bimodality in hub avPCC distributions do not appear to be robust across available %%@
interaction and expression data sets, and tests for the differences observed on deletion of the two hub types %%@
have not considered important outlier effects. Moreover, there is considerable evidence to suggest 
that the observed date/party distinction is at least partly an artefact, or consequence, of the different properties of the Y2H %%@
and AP/MS data sets.

In order to study the topological properties of hub nodes in greater detail, we partitioned protein interaction %%@
networks into communities and examined the statistics of the distributions of hub links. Our results show that %%@
hubs can exhibit an entire spectrum of structural roles and that, from this perspective, there is little evidence to suggest a %%@
definitive date/party classification. We find, moreover, that expression avPCC of a hub with its partners is not a %%@
strong predictor of its topological role, and that the extent of interacting protein co-expression varies considerably %%@
across the data sets that we examined. 

Additionally, a key issue with existing interaction networks is that they are incomplete.  We have compared %%@
some of the available `high-quality' yeast data sets and shown that they have very little overlap with each %%@
other. One can obtain protein interaction data using several experimental techniques, and each method appears %%@
to preferentially pick up different types of interactions \cite{vid08, lew09}. The only published interactome map 
of which we are aware that examines proteins in their natural cellular environment \cite{tar08} is largely 
disjoint with other data sets and shows little evidence of a date/party %%@
dichotomy. We find similar issues in human interaction data sets. A general conclusion about interactome %%@
properties is thus difficult to reach, as it would require robust results for a number of different species, %%@
which are unattainable at present due to the limited quantity and questionable quality of protein interaction %%@
and expression data.

As an alternative way of defining roles in the interactome, we have also investigated a link-centric approach, %%@
in which we study the topological properties of links (interactions) as opposed to nodes (proteins).  In %%@
particular, we examined link betweenness centrality as an indicator of a link's importance to network %%@
connectivity. We found that this too does not correlate significantly with expression PCC of the interacting %%@
proteins. For certain data sets, however, it does appear to correlate quite strongly with the functional %%@
similarity of the proteins.  Additionally, there appears to be a threshold value of betweenness centrality %%@
beyond which one observes a sudden drop in functional similarity. We also found that the high-betweenness %%@
interactions are enriched for kinase bindings and other kinds of interactions involved in signalling and %%@
transportation functions.  This suggests that a notion of intramodular versus intermodular interactions, %%@
somewhat analogous to the weak/strong tie dichotomy in social networks, might be more useful.  However, further %%@
work would be required to establish such a framework of elementary biological roles in protein interaction %%@
networks. As the quantity, quality, and diversity of protein interaction and expression data sets increases, we %%@
hope that this perspective will enhance understanding of the organisational principles of the interactome.

% You may title this section "Methods" or "Models". 
% "Models" is not a valid title for PLoS ONE authors. However, PLoS ONE
% authors may use "Analysis" 
\section*{Materials and Methods}

\subsection*{Protein Interaction Data Sets}

Several experimental methods can be used to gather protein interaction data. These include 
high-throughput yeast two-hybrid (Y2H) screening \cite{uet00, ito01, fro97, fro00}; affinity purification of %%@
tagged proteins followed by mass spectrometry (AP/MS) to identify associated proteins
\cite{ho02, gav02}; curation of individual protein complexes reported in the literature \cite{mew02}; and in %%@
silico predictions based on multiple kinds of gene data \cite{von02}. There is also a more recent
technique, known as the protein-fragment complementation assay (PCA) \cite{tar08}, which is able to detect %%@
protein-protein interactions in their natural environment within the cell. However, only one
large-scale study has used this technique thus far \cite{tar08}. Each of these methods gives an incomplete %%@
picture of the interactome; for instance, a recent aggregation of high-quality Y2H data sets for {\it Saccharomyces 
cerevisiae} %%@
(the best-studied organism) was estimated to represent only about 20\% of the whole yeast binary protein %%@
interaction network \cite{vid08}. 

Each technique also suffers from particular biases. It has been suggested that Y2H is likely to report binary interactions %%@
more accurately, and (due to the multiple washing steps involved in affinity purification) it is also expected to be better at detecting %%@
weak or transient interactions \cite{vid08}. Converting protein complex data into %%@
interaction data is also an issue with AP/MS. This method entails using a `bait' protein to `capture' other proteins that subsequently bind to it to form complexes. Once %%@
one has obtained these complexes and identified their proteins using mass spectrometry, one can assign %%@
protein-protein interactions using either the spoke or the matrix model \cite{hak07}. The spoke model only %%@
counts interactions between the bait and each of the proteins captured by it, whereas the matrix model counts %%@
all possible pairwise interactions in the complex. Unsurprisingly, the actual topology of the complex is %%@
generally different 
from either of these representations. On the other hand, AP/MS is expected to be more reliable at finding %%@
permanent associations. Two-hybrid approaches also do not seem to be particularly suitable for characterising %%@
protein complexes, giving rise to the view that complex formation is not merely the superposition 
of binary interactions \cite{gav02}. Thus, the two major techniques appear to be disjoint and to cover %%@
different aspects
of the interactome, and the differences between data sets from these sources perhaps correspond mostly to false %%@
negatives rather than false positives \cite{vid08}.

Given these factors, choosing which data sets to use for building and analysing the %%@
network is itself a significant issue (see the discussion in the main text). For our analysis, we chose to work %%@
predominantly with networks consisting of multiply-verified interactions, which are constructed from evidence %%@
attained using at least two distinct sources. Such data sets 
are unlikely to contain many false positives, but might include many false negatives (i.e., missing %%@
interactions). In Table \ref{tab:ds}, we summarise the data sets that we employed. Here are additional details %%@
about how they were compiled:

\begin{itemize}
\item Online Predicted Human Interaction Database (OPHID): This data was sent to us by Taylor {\it et al.} \cite{tay09};
it is an updated version of the interaction data used in their paper. It is based on their curation of the %%@
online OPHID repository \cite{bro05}; they have mapped proteins to their corresponding NCBI (National Center %%@
for Biotechnology Information) gene IDs. Additionally, we removed genes that did not have expression data in %%@
GeneAtlas \cite{su04} (avPCC cannot be calculated for these, as GeneAtlas is the only expression data set used by 
Taylor {\it et al.} \cite{tay09}), leaving a network with 8199 human gene IDs and 37968 interactions between them.

\item Filtered Yeast Interactome (FYI): Compiled by Han {\it et al.} \cite{vid04}. This was created by intersecting %%@
data generated by several methods, including Y2H, AP/MS, literature curation, in silico predictions, and the %%@
MIPS (\emph{http://mips.gsf.de/}) physical interactions list.  It contains 1379 proteins and 2493 %%@
interactions that were observed by at least two different methods.

\item Filtered High-Confidence (FHC): This data set was generated by Bertin {\it et al.} \cite{vid07} by filtering a %%@
data set called high-confidence (HC), which was compiled by Batada {\it et al.} \cite{bat06}. To conduct the %%@
filtration Bertin {\it et al.} applied criteria similar to those used for FYI and obtained 5991 %%@
independently-verified interactions amongst 2559 proteins. HC consists of 9258 interactions amongst 2998 %%@
proteins, taken from (published) literature-curated and high-throughput data sets, and they were also supposed to %%@
be multi-validated. However, Bertin {\it et al.} \cite{vid07} claimed that many interactions in HC had in fact been %%@
derived from a single experiment that was reported in multiple publications and thus removed such instances %%@
from it to generate FHC.

\item Database of Interacting Proteins core (DIPc): We obtained this data set from the DIP website
(\emph{http://dip.doe-mbi.ucla.edu/}). DIP is a large database of protein interactions compiled from a number %%@
of sources. The `core' subset of DIP consists of only the most reliable interactions, as judged manually by %%@
expert curators and also automatically using computational approaches \cite{dea02}. We used the version dated 7 %%@
October 2007, which contains 2808 proteins and 6212 interactions.

\item Protein-fragment Complementation Assay (PCA): This new experimental technique was used by Tarassov %%@
{\it et al.} \cite{tar08} to obtain an in vivo map of the yeast interactome that consists of 1124 proteins and 2770 %%@
interactions. An attractive feature of this data set is that it measures interactions between proteins in their %%@
natural cellular context, in contrast to other prominent methods, such as Y2H (which requires transportation to the cell %%@
nucleus) and AP/MS (which requires multiple rounds of in vitro purification). To our knowledge, this is the %%@
only published large-scale interaction study of this sort.

\item Center for Cancer Systems Biology Human Interactome version 1 (CCSB-HI1): This data set was constructed %%@
by Rual {\it et al.} \cite{vid05} using a high-throughput yeast two-hybrid system, which they employed to test %%@
pairwise interactions amongst the products of about 8100 human open reading frames. The data set, which contains  %%@
2611 interactions amongst 1549 proteins, achieved a verification rate of 78\% in an independent co-affinity %%@
purification assay (that is, from a representative sample of interactions in the data set, 78\% could be detected %%@
in the independent experiment). 
\end{itemize}

\subsection*{Betweenness Centrality}

Betweenness centrality is a way of quantifying the importance of individual nodes or links to the connectivity %%@
of a network. It is based on the notion of information flow in the network. The (geodesic) betweenness %%@
centrality of a node/link is defined as the number of pairwise shortest paths in the network that pass through %%@
that object \cite{fre77, gir02}. If there are multiple shortest paths between a pair of nodes, each one is %%@
given equal weight so that all of their weights sum to unity. Thus, the weighted count of all pairwise shortest %%@
paths passing through a given node/link equals its betweenness centrality.

For finite, sparse, unweighted networks such as the ones we study, one observes an interesting effect in %%@
the distribution of link betweenness centrality values. The distribution is almost normal, with the exception %%@
of a
large spike at a value well above the mean (see the long vertical bar of points in the plots
in Figure \ref{fig:simbce}). This results from the large number of nodes with degree 1.  The link that connects %%@
such a node to the rest of the network must have a betweenness of $N - 1$, where $N$ is
the total number of nodes in the network. Simply, this link must lie on the $N-1$ shortest paths that connect %%@
the
degree 1 node to all of the other nodes, and it cannot lie on any other shortest paths. Thus, for our networks, 
the link betweenness centrality distribution shows a strong spike at a value of precisely $N-1$.

\subsection*{Topological Metrics and Node Roles}

The \textit{within-community degree} refers to the number of connections a node has within its own community. %%@
It is normalised here to a $z$-score, which for the $i^{th}$ node is given by the formula
\begin{equation}
	z_i = \frac{\kappa_i - \bar{\kappa}_{s_i}}{\sigma_{\kappa_{s_i}}}\,,
\end{equation}
where $s_i$ denotes the community label of node $i$, $\kappa_i$ is the number of links of node $i$ to other %%@
nodes in the same community $s_i$, the quantity $\bar{\kappa}_{s_i}$ is the average of $\kappa_i$ for all %%@
nodes in community $s_i$, and $\sigma_{\kappa_{s_i}}$  is the standard deviation of $\kappa_i$ in community $s_i$. The %%@
\textit{participation coefficient} of node $i$ measures how its links are distributed amongst different %%@
communities. It is defined as \cite{gui05}
\begin{equation}
	P_i = 1 - \sum_{s=1}^{N} \left( \frac{\kappa_{is}}{k_i} \right)^2\,,
\end{equation}
where $N$ is the number of communities, $\kappa_{is}$ is the number of links of node $i$ to nodes in community %%@
$s$, and $k_i$ is the total degree of node $i$. The participation coefficient approaches $1$ if the links of %%@
node $i$ are uniformly distributed amongst all communities (including its own) and is $0$ if they are all %%@
within its own community.

In the main text, we plot all nodes in the network in a two-dimensional space using coordinates determined by %%@
within-community degree and participation coefficient, and we divide the space into regions that correspond to %%@
different node roles. The boundaries between regions are of course arbitrary, so for simplicity we have %%@
used the demarcations employed by Guimer\`{a} and Amaral \cite{gui05}. First, it is important to distinguish between 
`community hubs' and `non-hubs'; the former are defined as those nodes with within-community degree $z \ge 2.5$. 
In this context, the term `hub' is applied to nodes with high within-community degree \cite{gui05}, so `non-hubs' might have high
overall degree. One can further partition both `community hubs' and %%@
`non-hubs' on the basis of the participation coefficient $P$ as follows \cite{gui05}:
\begin{itemize}
\item Non-hubs can be divided into ultra-peripheral nodes ($P \le 0.05$---virtually all links within their %%@
own community), peripheral nodes ($0.05 < P \le 0.62$---most links within their own community), non-hub %%@
connector nodes ($0.62 < P \le 0.80$---links to many other communities), and non-hub kinless nodes ($P > %%@
0.80$---links distributed roughly homogeneously amongst all communities).
\item Community hubs can be divided into provincial hubs ($P \le 0.30$---vast majority of links within own %%@
community), connector hubs ($0.30 < P \le 0.75$---many links to most other communities), and kinless hubs ($P %%@
> 0.75$---links distributed roughly homogeneously amongst all communities).
\end{itemize} 
We depict these 7 roles as demarcated regions in the plots in Figure \ref{fig:fcpcc}.

\subsection*{Jaccard Distance}

If one has two partitions of a given set of nodes, and a node $i$ is part of subset (or community) $C_i^1$ of nodes %%@
in one partition and part of subset $C_i^2$ in the other partition, then the Jaccard distance \cite{jac01} for node %%@
$i$ across the two partitions is defined as 
\begin{equation}
	J(i) = 1 - |C_i^1 \cap C_i^2|/|C_i^1 \cup C_i^2|\,.
\end{equation}
The symbols $\cap$ and $\cup$ correspond, respectively, to set intersection and union, and $|C|$ denotes the %%@
number of elements in set $C$. A Jaccard distance of 0 corresponds to identical communities, whereas the distance approaches 1
for very different communities. By averaging $J(i)$ over all nodes in the set, we can get an estimate %%@
of the similarity of the two partitions.

\subsection*{Functional Similarity}

In order to compute the functional similarity of two interacting proteins, we first define the set information
content (SIC) \cite{res95} of each term in our ontology for a given data set.
Suppose the complete set of proteins is denoted by $S$, and the subset annotated by term $i$ is denoted
by $S_i$.  The SIC of the term $i$ is then defined as
\begin{equation}
	SIC(i) = -\log_{10} \left( \frac{|S_i|}{|S|} \right)\,.
\end{equation}
Now suppose that we have two interacting proteins called $A$ and $B$. Let $S_A$ and $S_B$, respectively, denote %%@
their complete sets of annotations (consisting of not only their leaf terms but also all of their ancestors) %%@
from the ontology.  Then the functional similarity of the proteins is given by 
\begin{equation}
	f(A,B) = \frac{\displaystyle \sum_{i \in (S_A \cap S_B)} SIC(i)}{\displaystyle \sum\limits_{j \in (S_A \cup %%@
S_B)} SIC(j)}\,.
\end{equation}

% Do NOT remove this, even if you are not including acknowledgments
\section*{Acknowledgments}
We thank Patrick Kemmeren for providing the yeast expression data sets; Nicolas Simonis for giving details of
date/party hub classification; Jeffrey Wrana, Ian Taylor and Katherine Huang for the curated OPHID %%@
human interaction data, and GeneAtlas expression data; Pao-Yang Chen and Waqar Ali for pointing us to relevant online %%@
data repositories; Peter Mucha and Stephen Reid for useful discussions and help with some code; Anna Lewis %%@
for several useful discussions, as well as providing GO annotation 
data and MATLAB code for computing functional similarities; and George Nicholson and Max Little for useful %%@
discussions.

%\section*{References}
% The bibtex filename
\bibliography{biblio}

%\section*{Figures}
%\begin{figure}[!ht]
%\begin{center}
%%\includegraphics[width=4in]{figure_name.2.eps}
%\end{center}
%\caption{
%{\bf Bold the first sentence.}  Rest of figure 2  caption.  Caption 
%should be left justified, as specified by the options to the caption 
%package.
%}
%\label{Figure_label}
%\end{figure}

%\section*{Tables}
%\begin{table}[!ht]
%\caption{
%\bf{Table title}}
%\begin{tabular}{|c|c|c|}
%table information
%\end{tabular}
%\begin{flushleft}Table caption
%\end{flushleft}
%\label{tab:label}
% \end{table}

\appendix

\section{Supplementary Text S1: Communities in the Interactome}

A network consists of elements (called nodes) that are connected to each other by edges (called links). Many %%@
real-world networks can be %%@
divided naturally into close-knit sub-networks called communities.  The investigation of algorithms for %%@
detecting communities in networks has received considerable attention in recent years \cite{for10,por09}.
 
From an intuitive standpoint, communities should consist of groups of nodes such that there are many links %%@
between nodes in the same group but few links between nodes in different groups. To detect communities %%@
algorithmically, this notion must be formalised. In order to identify community structure in the various interaction networks that we
examine, we employ a method based on optimising the well-known quality function known as graph `modularity' %%@
\cite{new04, new06}. Suppose that an %%@
unweighted network with $n$ nodes and $m$ links is divided into $N$ communities $C_1, C_2, \cdots, C_N$. Let %%@
$k_i$ denote the degree (number of links) of node $i$ and let $A$ be the $n \times n$ adjacency matrix, so that %%@
$A(i,j)$ is $1$ if nodes $i$ and $j$ have a link between them and $0$ if they do not. The modularity $Q$ is %%@
then given by \cite{new06}
\begin{equation}
	Q = \frac{1}{2m} \sum_{l=1}^N \sum_{i,j \in C_l} \left(A_{ij} - \frac{k_i k_j}{2m}\right)\,, \label{modmod}
\end{equation}   
where $k_i k_j/(2m)$ is the expected number of links between nodes $i$ and $j$ in a network with the same %%@
expected degree distribution but with links placed at random. Graph modularity thus captures how
many more links there are within the specified communities than one would expect to see by chance in a network %%@
with no modular structure. Note, however, that (\ref{modmod}) assumes a particular null model that explicitly %%@
preserves the expected degree distribution in the random setting. It is possible to employ other null models %%@
\cite{por09}, though this is the most common choice. In fact, we use an extension of this method based on an analogy
to the Potts model in statistical mechanics \cite{rei06}. This incorporates a resolution parameter (denoted by $\gamma$)
into the equation for modularity, leading to the quality function
\begin{equation}
	H = \frac{1}{2m} \sum_{l=1}^N \sum_{i,j \in C_l} \left(A_{ij} - \gamma \frac{k_i k_j}{2m}\right)\,. \label{potts}
\end{equation}
Setting $\gamma=1$ leads to the standard modularity function (\ref{modmod}), which is what we use for the results in 
Figure 3. However, we also present results for $\gamma=0.5$ and $\gamma=2$ in Figure S2, demonstrating that whilst the 
number of communities changes substantially as we increase or decrease the resolution, the pattern of role assignments to the
nodes remains similar.

Using this framework, we can detect communities by maximising the quality function (\ref{potts}) over all possible network %%@
partitions. Because this problem is known to be NP-complete \cite{bra08}, reliably finding the global maximum is %%@
computationally intractable even for small networks. Thankfully, there exist a number of good computational %%@
heuristics that can be used to obtain good local maxima \cite{dan05,for10,por09}. Here we use recursive %%@
spectral bisection \cite{new206}. One can use similar procedures to optimise quality functions other than %%@
modularity, and the analysis that we have described can be employed in those cases as well.

Maximising graph modularity (\ref{modmod}) is expected to give a partition in which the density of links %%@
within each community is significantly higher than the density of links between communities. In Figure %%@
S4, we show the network partition (with nodes coloured according to community) that results from %%@
applying such an optimisation to the largest connected component of the filtered yeast interactome (FYI) data set \cite{vid04}.  

To assess how well the obtained topological communities reflect functional organisation, we used %%@
annotations from the Gene Ontology (GO) database \cite{ash00} to define their {\it Information Content} ($IC$). 
GO provides a controlled vocabulary for describing genes and gene products, %%@
such as proteins, using a limited set of annotation terms. It consists of three separate ontologies---one 
each for biological process, cellular component, and molecular function. For each community, we computed %%@
the $p$-value of the most-enriched GO annotation term; the frequency of this term within its community is %%@
highest relative to its background frequency in the entire network. To do this, we used the hypergeometric distribution, which %%@
corresponds to random sampling without replacement. The extent of enrichment can then be gauged using $IC$ \cite{res95}, 
which is defined as 
\begin{equation}
	IC = -\log_{10}(p)\,,  \label{ic}
\end{equation}
where $p$ denotes the $p$-value. In Table S1, we summarise the results of calculating this measure for 
communities detected (for $\gamma=1$) on two of the yeast interaction data sets, FYI and the more recent filtered high-confidence (FHC \cite{vid07}). 
For comparison, we also examine a random partition of FYI into communities %%@
with the same size distribution as the actual ones.

It is clear that on average there is very significant functional enrichment within the 
detected communities.  In particular, it is far greater than could be expected by chance. This is in accordance %%@
with previous studies on communities in protein interaction networks \cite{dun05, ada06, che06, mar08, lew09}. %%@
It is also evident that $IC$ varies widely over communities and that not all of them are equally enriched. There %%@
are some relatively heterogeneous communities (which are not aptly described by a single, specific GO term) and %%@
others that show a very high functional coherence. In particular, a more detailed inspection of the community %%@
composition reveals that proteins that are part 
of the large and small ribosomal subunit complexes are almost perfectly grouped together, and several other %%@
communities consist exclusively of proteins that are known to be part of a given complex. Thus, the topology of %%@
the interaction network provides a great deal of information about the functional organisation of the proteome. %%@
We do not claim that our particular partitioning is in any sense unique; rather, it is only a means to an end, %%@
as our aim is to examine the implications of community structure for individual protein roles, with particular %%@
reference to the notion of date and party hubs. We have also used the greedy algorithm described by Blondel {\it et al.} \cite{blo08} 
as an alternative method for optimising modularity. We obtained results that are very similar to those presented here.

\makeatletter \renewcommand{\thefigure}{S\@arabic\c@figure} \makeatother
\setcounter{figure}{0}

\begin{figure}[!ht]
\begin{center}
\subfigure[]{\scalebox{0.5}{\includegraphics{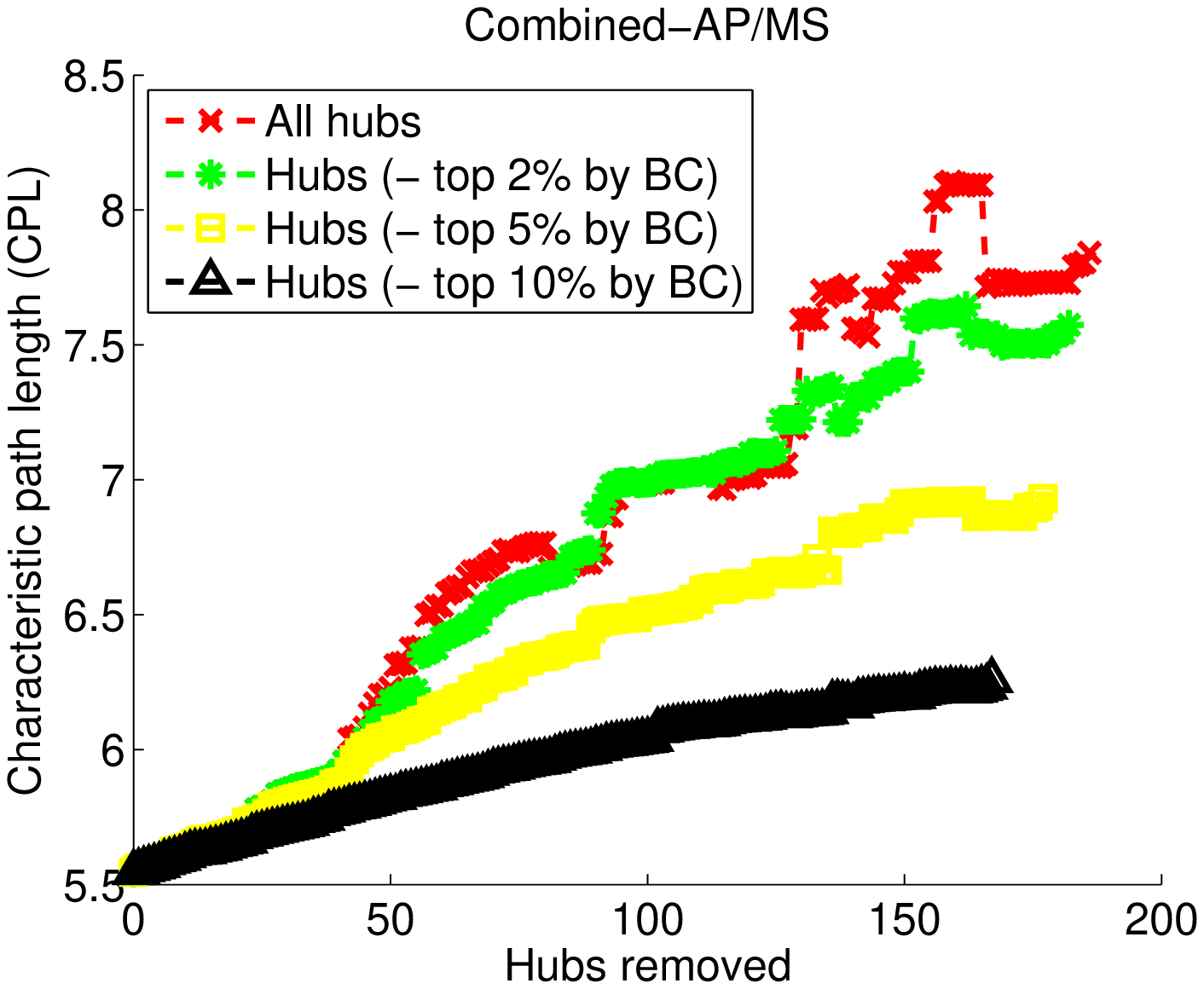}}}
\subfigure[]{\scalebox{0.5}{\includegraphics{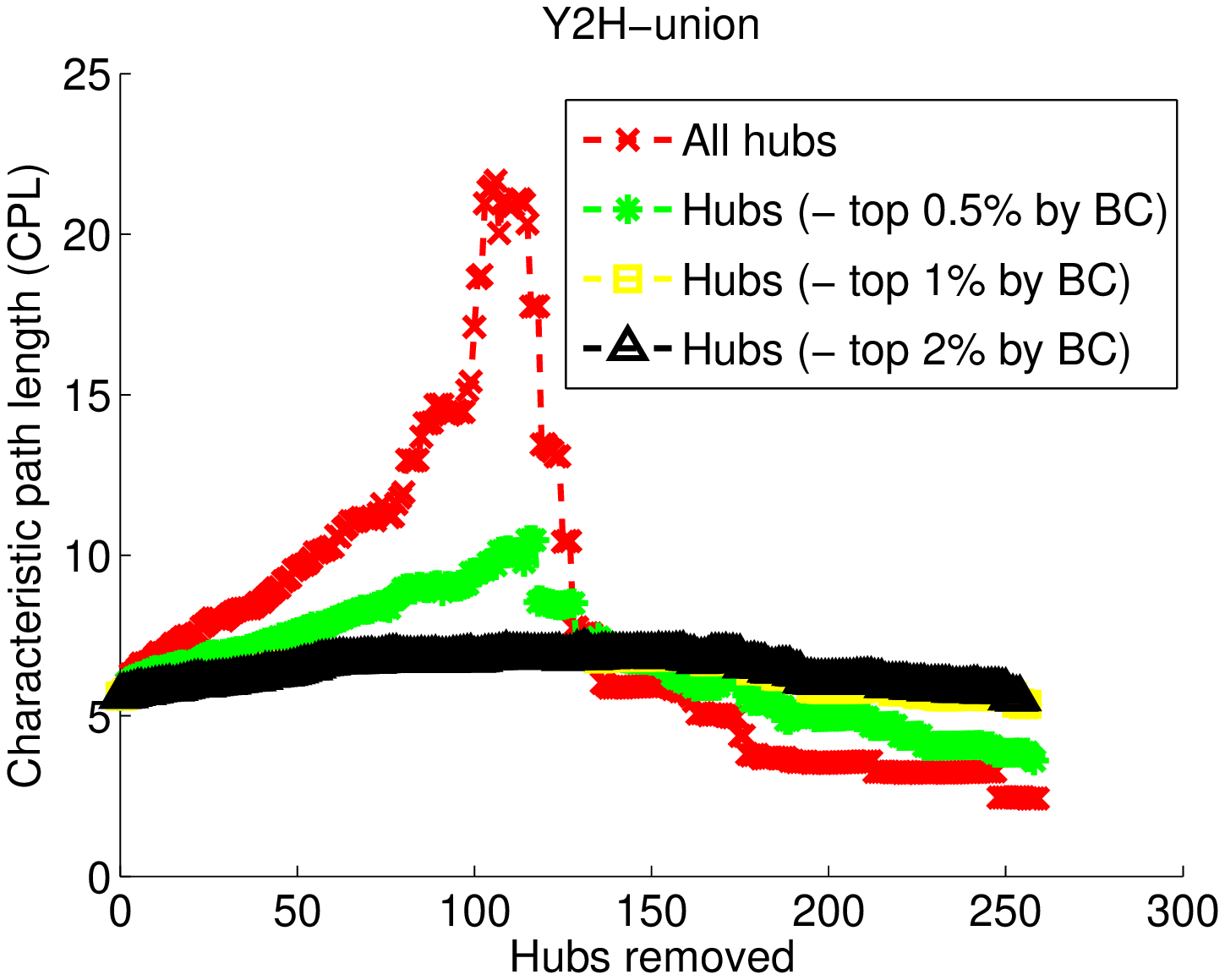}}}
\caption{{\bf Hub deletion effects for AP/MS-only and Y2H-only data sets.} Change in characteristic path length (CPL, the mean length of all 
finite pairwise shortest paths) on removal of hubs in decreasing order of degree
from the (a) `Combined-AP/MS' and (b) `Y2H-union' data sets \cite{vid08}. The `top X\%' captions refer to deletion of all hubs except the X\% with the highest 
betweenness centrality values. Note that when deleting the full sets of hubs, the Y2H network shows a much more dramatic increase in CPL, which may suggest 
that date hubs are more crucial to network connectivity than party hubs (the Y2H hubs being predominantly date hubs, whereas the AP/MS hubs are mostly party 
hubs \cite{vid08}). However, only a very tiny fraction of Y2H-union hubs seem to be responsible for the huge CPL increase on deletion, and protecting
these few high-betweenness hubs greatly reduces the impact of hub deletion on network connectivity. This shows that the vast majority of those referred to 
as date hubs are on average no more central to the network than the party hubs.} 
\label{fig:delapy}
\end{center}
\end{figure}

\begin{figure}[!ht]
\begin{center}
\subfigure[FHC ($\gamma=0.5$, 35 communities)]{\scalebox{0.5}{\includegraphics{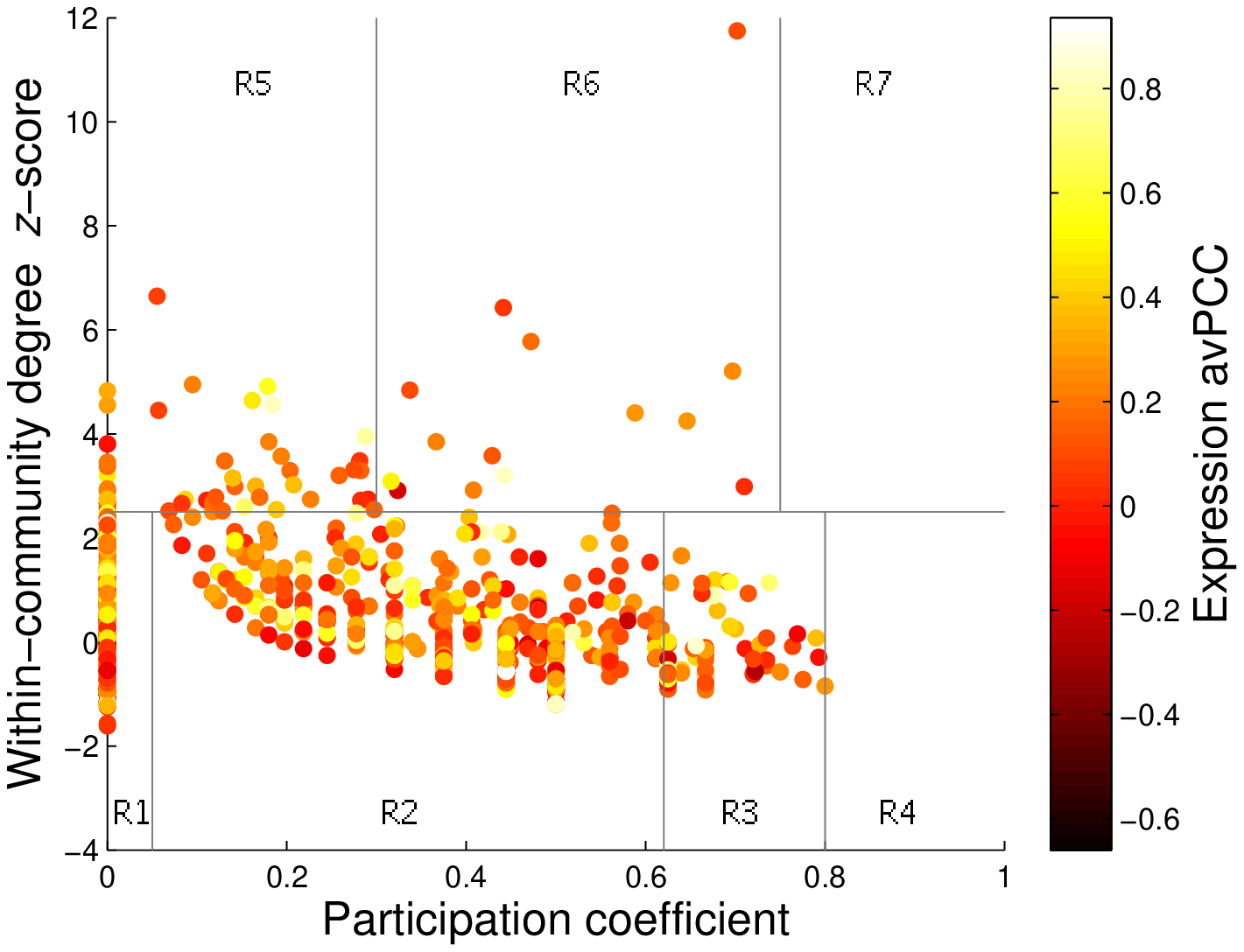}}}
\subfigure[CCSB-HI1 ($\gamma=0.5$, 32 communities)]{\scalebox{0.5}{\includegraphics{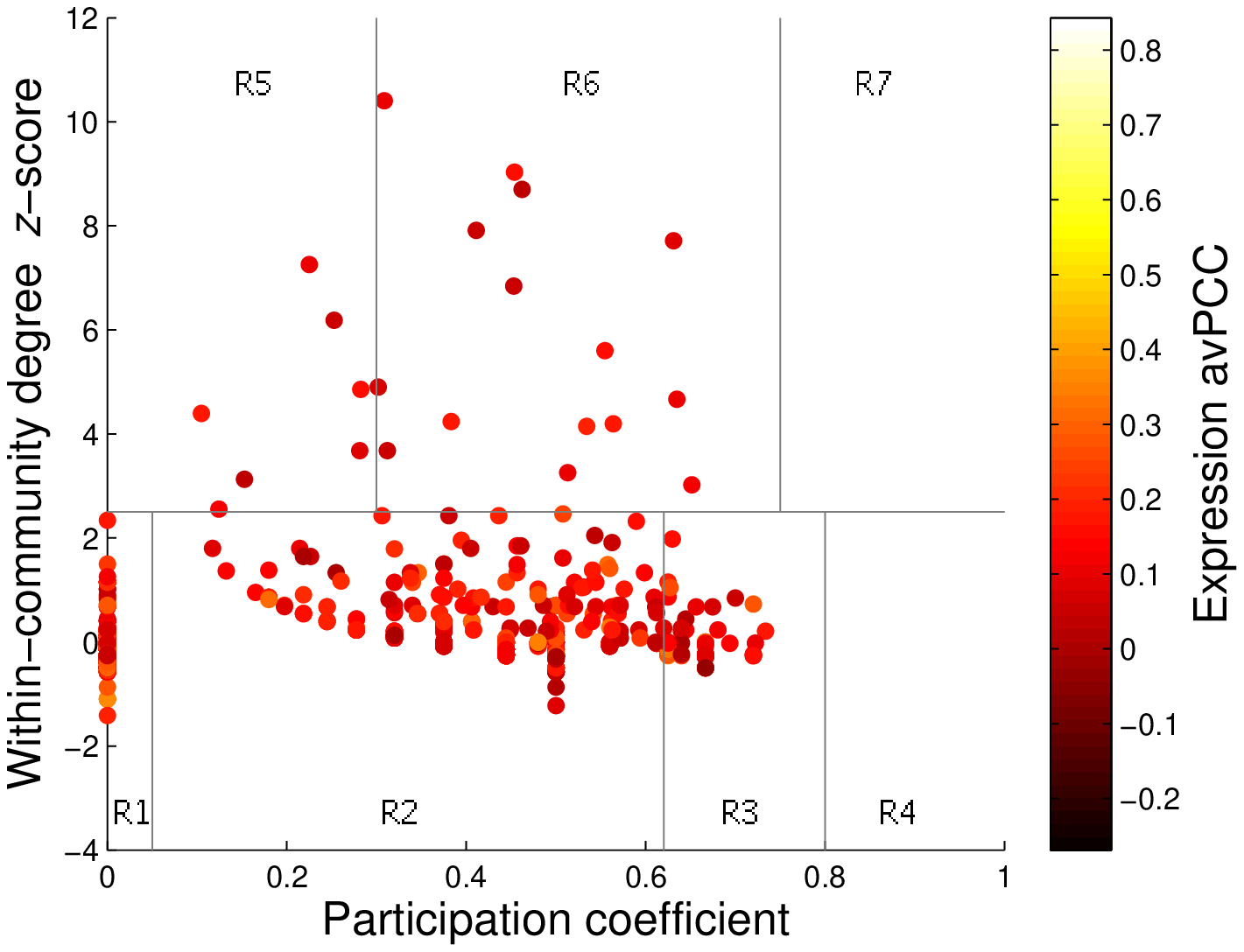}}}
\subfigure[FHC ($\gamma=2$, 81 communities)]{\scalebox{0.5}{\includegraphics{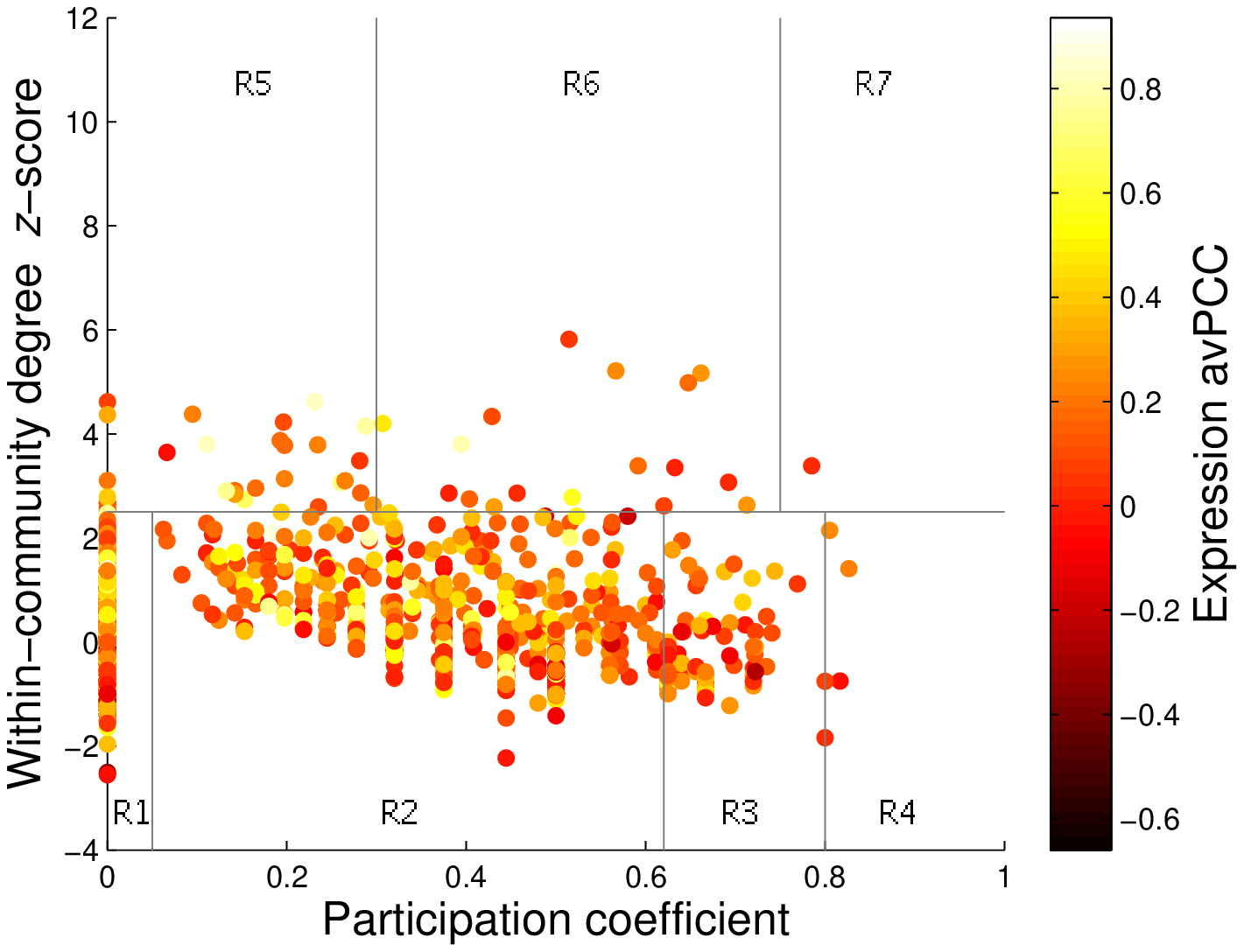}}}
\subfigure[CCSB-HI1 ($\gamma=2$, 71 communities)]{\scalebox{0.5}{\includegraphics{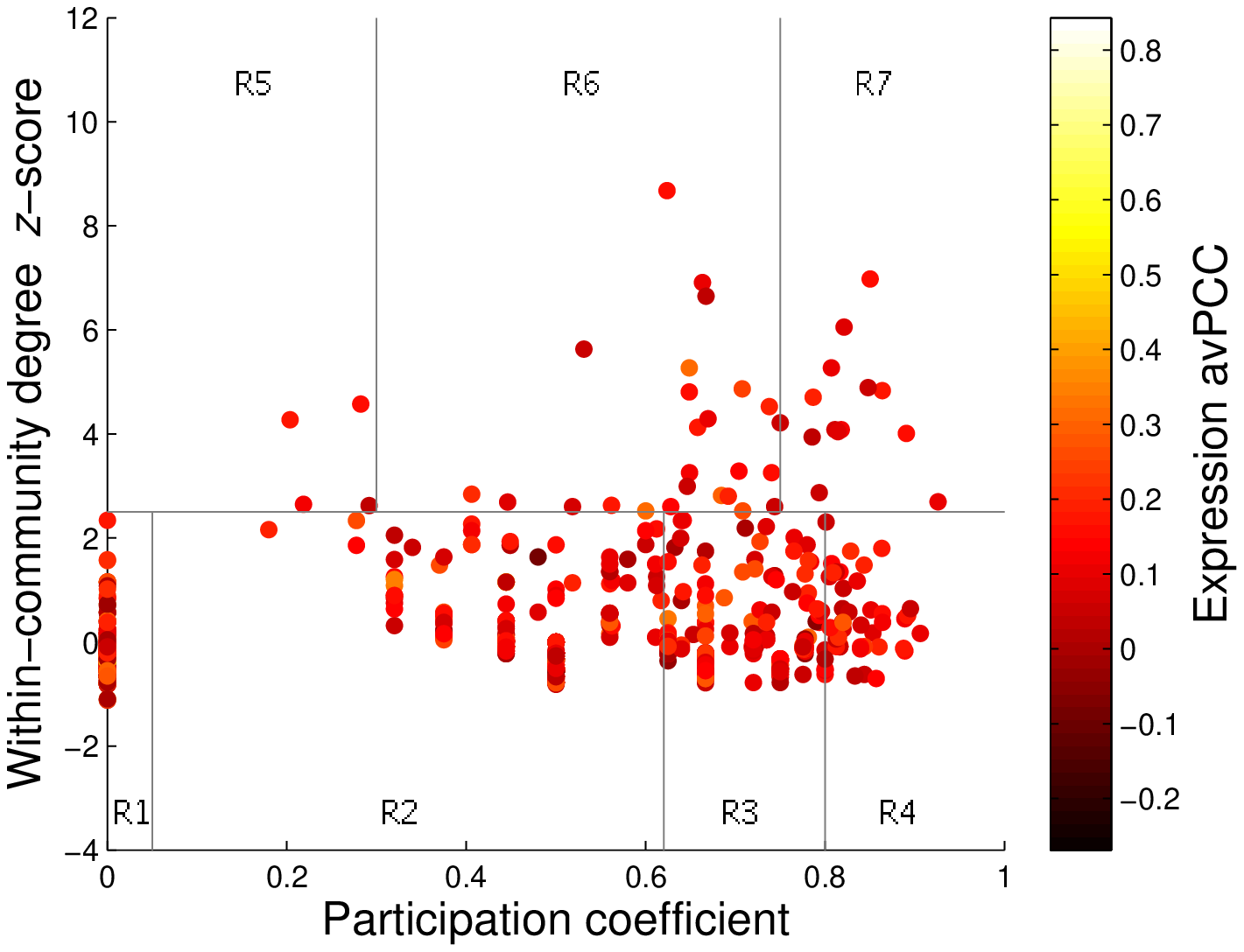}}}
\caption{{\bf Topological node role assignments and relation with avPCC.} Plots for (a),(c) Yeast network 
(FHC \cite{vid07}---2,233 nodes) and (b),(d) Human network (CCSB-HI1 \cite{vid05}---1,307 nodes). 
As in Figure 3, but with different resolution parameter ($\gamma$) settings used for community detection (see Text S1).
}

\label{fig:fcpccs}
\end{center}
\end{figure}

\begin{figure}[!ht]
\begin{center}
\subfigure[Biological Process]{\scalebox{0.6}{\includegraphics{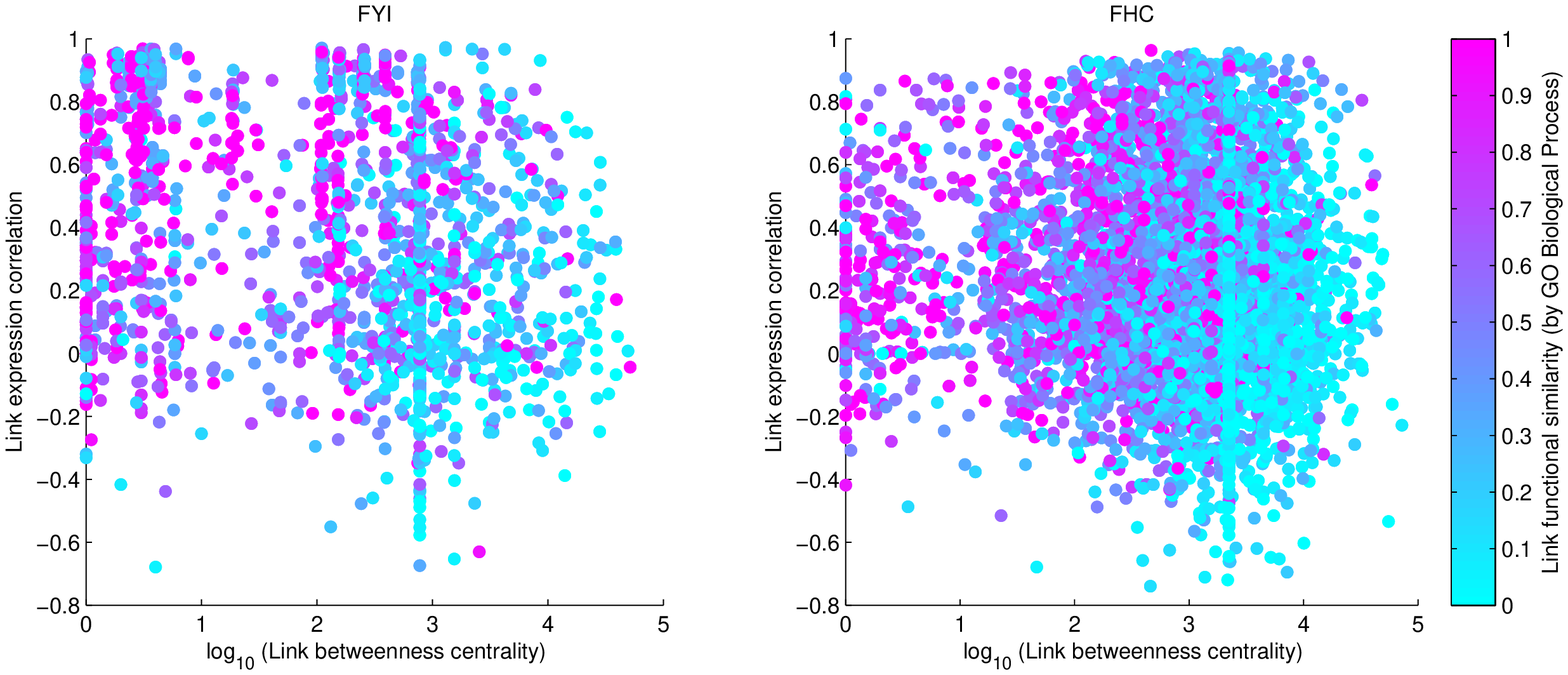}} \label{bp}}
\subfigure[Molecular Function]{\scalebox{0.6}{\includegraphics{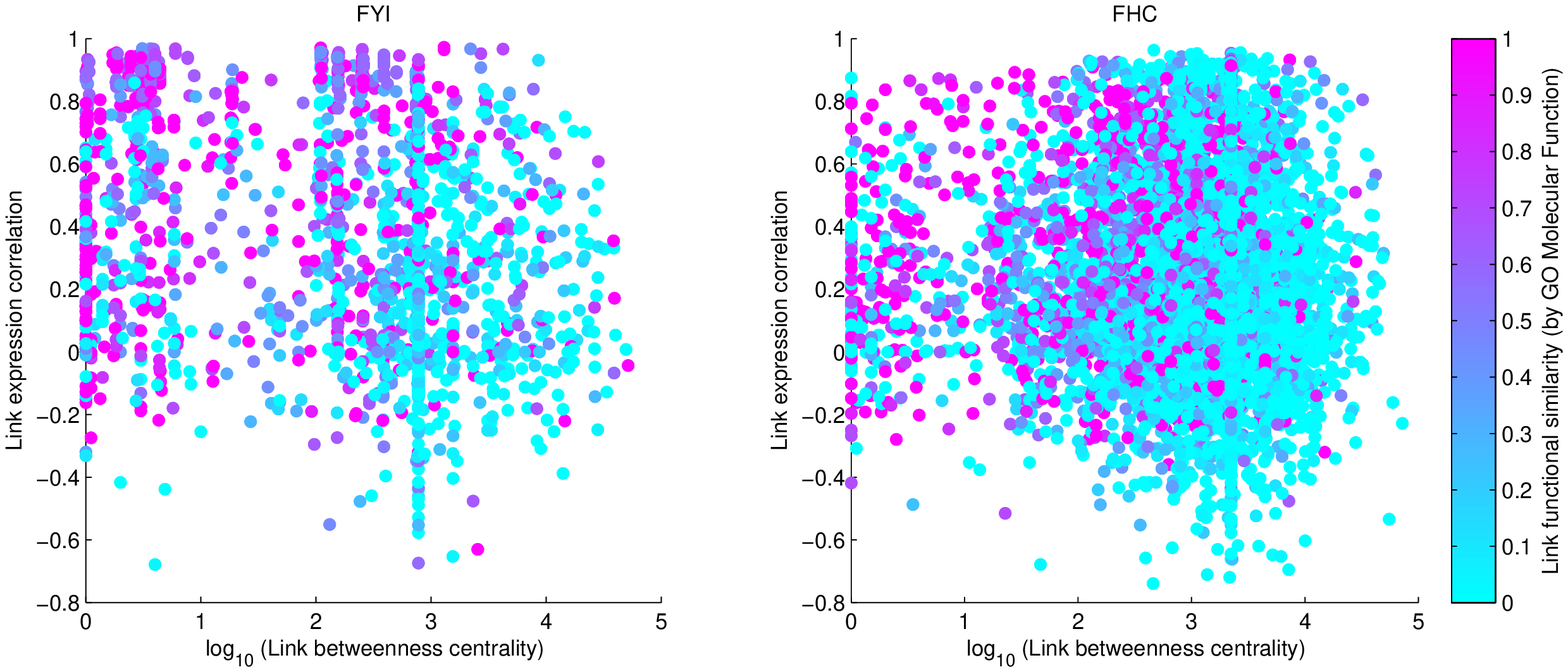}} \label{mf}}
\caption{{\bf Relating interaction betweenness, co-expression, and functional similarity.}
Plots show link betweenness centralities versus expression correlations, with points coloured according to average  
similarity of interactors' (a) GO (Biological Process) and (b) GO (Molecular Function) annotations, 
for two protein interaction data sets: FYI \cite{vid04} (778 nodes, 1,798 links) and FHC \cite{vid07} (2,233 nodes, 5,750 links).  
Pearson correlation coefficient values of log(link betweenness) with functional similarity are (a) $-0.41$ ($z$-score $\approx %%@
-18.6$, 
$p$-value $\approx 3.9 \times 10^{-77}$) for FYI, $-0.42$ ($z$-score $\approx -33.9$, $p$-value $\approx 4.7 \times 10^{-252}$) for %%@
FHC; 
(b) $-0.39$ ($z$-score $\approx -17.3$, $p$-value $\approx 4.5 \times 10^{-67}$) for FYI, $-0.31$ ($z$-score $\approx -24.7$, 
$p$-value $\approx 1.6 \times 10^{-134}$) for FHC.  
}

\label{fig:simbpmf}
\end{center}
\end{figure}

\begin{figure}[!ht]
\begin{center}
\scalebox{0.5}{\includegraphics{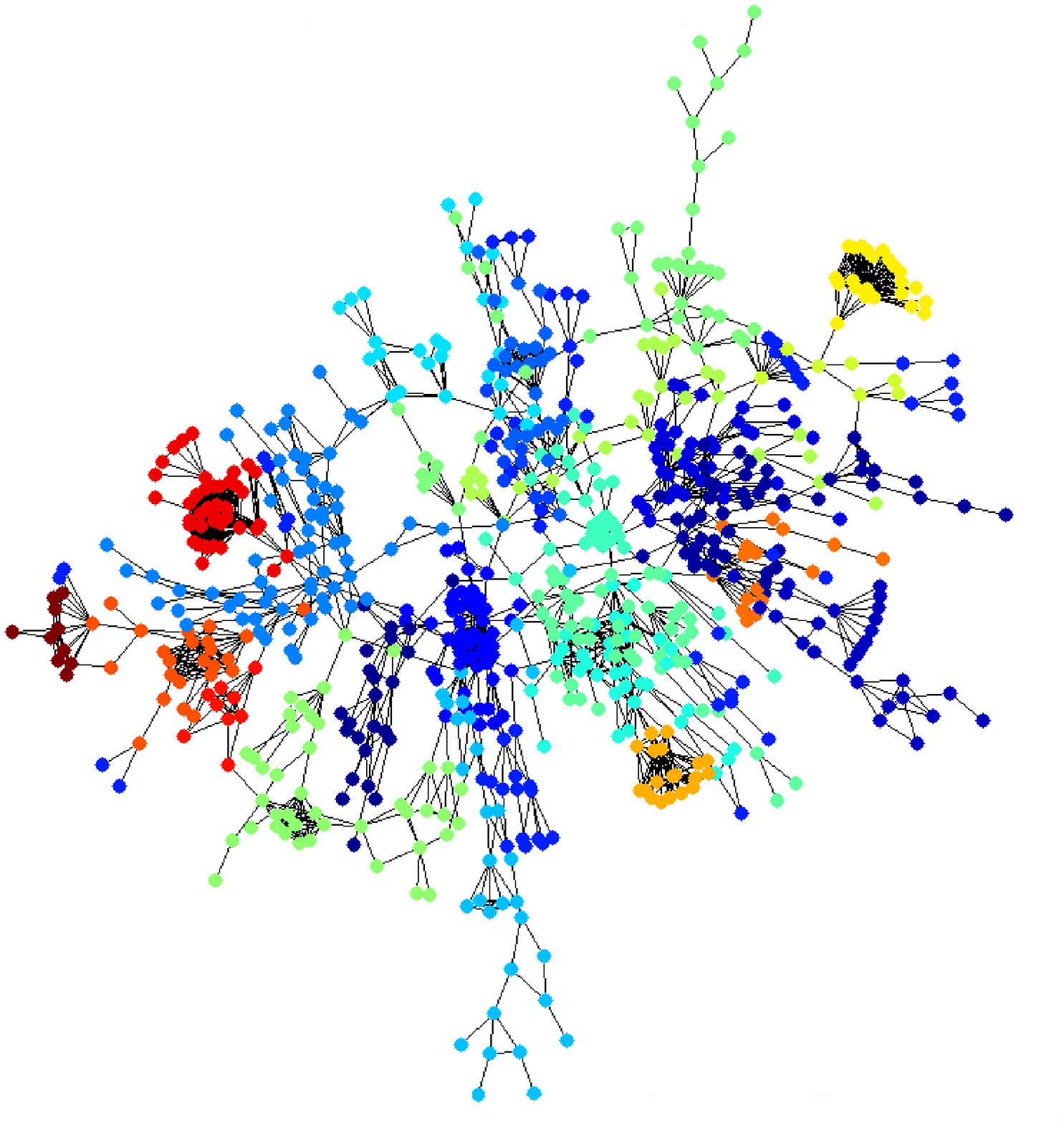}}
\caption{{\bf Community structure in the largest connected component of the FYI network \cite{vid04}.} The %%@
different colours correspond to different communities (25 in all). The graph modularity value for this %%@
partition is 0.8784. We generated this visualisation using the Kamada-Kawai algorithm \cite{kam89}.  
}

\label{fig:fyi}
\end{center}
\end{figure}

\makeatletter \renewcommand{\thetable}{S\@arabic\c@table} \makeatother
\setcounter{table}{0}

\begin{table}[!ht]
\begin{center}
\begin{tabular}{|c|c|c|c|c|c|c|c|c|c|c|c|c|c|}
\hline
\bf{Data} & \bf{Commu-} & \multicolumn{3}{c|}{\bf{MF $IC$}} & \multicolumn{3}{c|}{\bf{CC $IC$}} &
\multicolumn{3}{c|}{\bf{BP $IC$}} & \multicolumn{3}{c|}{\bf{Best $IC$}} \\
\bf{set} & \bf{nities} & Min&Max&Avg & Min&Max&Avg & Min&Max&Avg & Min&Max&Avg \\
\hline
FYI & 25 & 2.05 & 43.09 & 14.36 & 4.28 & 51.60 & 17.18 & 2.99 & 35.74 & 15.72 & 4.81 & 51.60 & 20.15 \\
FYI & 25 (random) & 1.28 & 2.78 & 1.88 & 1.25 & 3.00 & 2.07 & 1.46 & 3.04 & 2.13 & 1.46 & 3.04 & 2.36 \\
\hline
FHC & 63 & 1.47 & 51.37 & 11.22 & 0.11 & 68.18 & 16.40 & 1.73 & 98.51 & 17.08 & 1.97 & 98.51 & 20.08 \\  
\hline
\end{tabular}
\caption{{\bf Evaluating community partitions.} Information Content ($IC$) of the most enriched term for each of the three GO ontologies (MF -- %%@
Molecular Function; CC -- Cellular Component; BP -- Biological Process) and over all three ontologies combined %%@
(`Best $IC$'). We give the minimum, maximum, and average $IC$ over all of the communities (at the default resolution $\gamma=1$) that we detected in two %%@
data sets, FYI \cite{vid04} and FHC \cite{vid07}. We generated the random communities for FYI using the same-size distribution to the actual %%@
ones. In other words, we remove the actual community labels of all proteins and then randomly re-assign them (using one label %%@
per protein).}
\label{tab:te}
\end{center}
\end{table}

\end{document}